%% file: main.tex
\documentclass[12pt]{article} 
\usepackage[margin=1.5cm,nohead]{geometry}
\usepackage{graphicx}
\usepackage{amsmath}
\usepackage{amsthm}
\usepackage{amssymb}
\usepackage{sgame}
\usepackage{color}
\usepackage{multirow}
\usepackage{epsfig}
\usepackage{setspace}
\usepackage{nccmath}
\usepackage{xfrac}
\usepackage[]{subfig}
\usepackage{accents}
\usepackage{bbm}
\usepackage{tabularx}
\usepackage{algorithm}
\usepackage{algorithmic}
\usepackage{verbatim}
\usepackage{lscape}
\usepackage{afterpage}
\usepackage{rotating}
\usepackage[font=footnotesize,labelfont=bf]{caption}
\usepackage{fancyhdr}
\usepackage{hyperref}
\usepackage{afterpage}
\usepackage{caption}
\usepackage{longtable}
\usepackage{dcolumn}
\usepackage{threeparttable}

\begin{document}

\title{The Internet as Quantitative Social Science Platform: Insights From a Trillion Observations}

\author{%
	Klaus Ackermann\thanks{Center for Data Science and Public Policy, University of Chicago, USA; Department of Economics, Monash University, Australia} \and
	Simon D Angus\thanks{ Department of Economics, Monash University, Australia} \and
	\and
	Paul A Raschky\footnotemark[2]}

\maketitle

\begin{abstract}
{\noindent With the large-scale penetration of the internet, for the first time, humanity has become linked by a single, open, communications platform. Harnessing this fact, we report insights arising from a unified internet activity and location dataset of an unparalleled scope and accuracy drawn from over a trillion (1.5$\times 10^{12}$) observations of end-user internet connections, with temporal resolution of just 15min over 2006-2012. We first apply this dataset to the expansion of the internet itself over 1,647 urban agglomerations globally\cite{angel2010planet}. We find that unique IP per capita counts reach saturation at approximately one IP per three people, and take, on average, 16.1 years to achieve; eclipsing the estimated 100- and 60- year saturation times for steam-power and electrification respectively\cite{Jovanovic:2005vv}. Next, we use intra-diurnal internet activity features to up-scale traditional over-night sleep observations, producing the first global estimate of over-night sleep duration in 645 cities over 7 years\cite{Roenneberg:2013hc}. We find statistically significant variation between continental, national and regional sleep durations including some evidence of global sleep duration convergence. Finally, we estimate the relationship between internet concentration and economic outcomes in 411 OECD regions and find that the internet's expansion is associated with negative or positive productivity gains, depending strongly on sectoral considerations. To our knowledge, our study is the first of its kind to use online/offline activity of the entire internet to infer social science insights, demonstrating the unparalleled potential of the internet as a social data-science platform.}
\end{abstract}


\emph{Abbreviations:} IP, internet protocol; GPT, general purpose technology

\newpage
\section{Introduction}
By any measure, the internet -- the physical web of fibre-optic cabling connected to myriad routers, servers and devices -- is a most remarkable invention of humankind, affecting the full spectrum of human behaviour including health care\cite{vonMutius:2012cb,Hesse:2016by}, political activity\cite{Falck:2014hv}, time-use decisions\cite{Lam:2014eb}, and even the most intimate of human relationships\cite{Manning:2007ft}. By 2016, it has been estimated that 3.5 billion individuals (47.1\%) will be online, or almost 1 billion households (52.3\%), together accessing a global bandwidth capacity of 185,000 Gbit/s (around 10.1 billion pages/s), ten times that which was available in 2008\cite{ITU:2016tu}. These figures are all the more remarkable given that less than 5\% of the world's population was online at the turn of the millenium\cite{ITU:2009ww}.

Yet for the social scientist, these figures suggest a different kind of wonder -- for the first time in human history, half of the world's population is now connected to a \emph{single} general purpose technology (GPT)\cite{Bresnahan:1995im,Jovanovic:2005vv}. A desktop computer in an internet cafe in Nairobi, a data-enabled cell-phone in Melbourne, an iPad on the wireless network of a bullet train out of Kyoto -- each is assigned an Internet Protocol (IP) address on \emph{one} network. Furthermore, owing to the thoroughly democratic foundations of the internet, each device can instantly, and at negligible cost, \emph{passively} query the other's `online' or `offline' status (i.e. without perturbing the target device in any way). The global reach of the internet coupled to its passive message technology thus powerfully sets the internet apart as a \emph{social data-science platform}\cite{Giles:2012gq} from traditional data-gathering approaches.

However, the technical challenges involved in generating internet query data, at global scale, are immense. To exhaustively probe every public, allocated, IP address under the Internet Protocol version 4 (IPv4) addressing system (i.e. around half of the possible 4.3$\times 10^9$ addresses) can take between 24 and 70 days\cite{Heidemann:2008cv}, although representative activity scans can be accomplished in minutes\cite{Quan:2014uh}. Additionally, IP address blocks do not have stable geo-locations over time, hence, any spatial analysis must be able to recover, from additional, historical, IP/geo-location scans, the location of a given active or inactive IP address at a point in time\cite{Hu:2012hu}. Such features have created significant barriers for social scientists wishing to analyse global internet activity, resulting in internet subscription\cite{Clarke:2015ec} or internet infrastructure\cite{Falck:2014hv} data being used as common proxies.

\begin{figure*}[p]
\captionsetup{width=11.4cm}
\begin{center}
\centerline{\includegraphics[width=11.4cm]{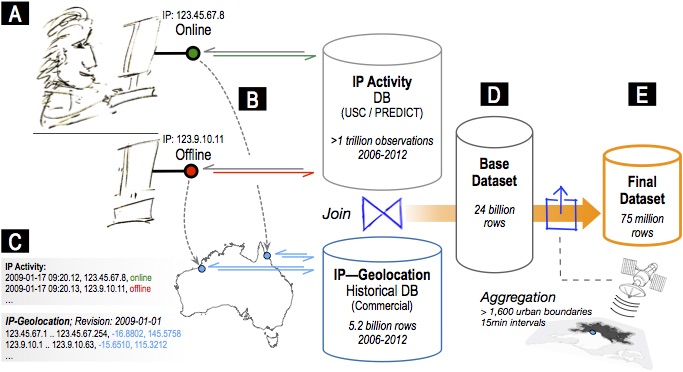}}
\caption{\textbf{Building the geo-located, Internet Protocol (IP) activity dataset.} \textbf{(a)} Every user on the internet is assigned a unique ID known as an IP address, a sequence of 4 integers in the range 0 to 255 (e.g. 123.45.67.8). The IP Activity DB is populated by periodically scanning every IP address. When a user has an open pathway to the internet it will respond as ‘online’ when an ICMP ‘probe’ is sent by the scan . Any non-response from the IP (e.g. the user’s modem is off or ‘asleep’, or a firewall is present) will be registered in the IP Activity DB as ‘offline’. Any IP address which can never be online at the time of the scan (i.e. the IP is missing from routing tables) is automatically discarded. \textbf{(b)} The geolocation (lon, lat) of an IP can be determined by repeated scanning from multiple remote locations. Since IPs are not statically assigned to users, an IP’s geolocation is updated every few weeks to form the IP-Geolocation historical DB. \textbf{(c)} Since the IP Activity DB provides {date-stamp, IP, activity} observations but the IP-Geolocation DB provides {IP-range, lon, lat} observations for a given revision, joining the two DBs requires nontrivial data-manipulation techniques on distributed hardware. \textbf{(d)} After the join, 24 billion geo-located, IP activity, observations resulted. \textbf{(e)} Finally, the base observations are spatially aggregated using over 1,600 urban boundaries obtained from satellite observational data, and temporally aggregated to 15min intervals, yielding a final dataset of 75 million rows.}\label{figure1}
\end{center}
\end{figure*}

In this report we demonstrate human behavioural insights arising from our group's successful joining of over a trillion (1.5$\times 10^{12}$) IP activity (`offline'/`online') observations obtained during 2006-2012 to a highly-accurate, commercially-available, IP-geolocation library (Fig.~\ref{figure1}). Our approach yields a refined dataset of unparalleled scope and granularity: 75 million rows of online/offline observations over temporal segments of as little as 15min, at over 1,600 urban-boundaries\cite{angel2010planet} (cities) within 122 nations.

To demonstrate the scientific potential of these data we present first a formal characterisation of the growth dynamics of the internet itself, followed by an estimation of global human sleep duration founded on diurnal internet activity, and finish with an exploration of the relationship between sub-national internet penetration and economic outcomes. To the best of our knowledge, each application is the first of its kind.

\section{Diffusion}

The diffusion of technology, including previous GPTs\cite{Bresnahan:1995im,Jovanovic:2005vv}, is of ongoing interest in the economic literature\cite{Griliches:1957uq,Dixon:1980tb,Niosi:2012ed}. Previous related studies have used a variety of internet penetration proxies at either snapshot- or annualised- detail, each proxy having one or more compromises such as data-quality problems (in the case of ITU surveys\cite{Clarke:2015ec,BenitezBaleato:2015hk}), or actual internet use identification complications (in the case of block-based or router-based assignment\cite{Milner:2006gi,Weidmann:2016jv}). In contrast, since we observe actual end-user IP connections, in well-defined urban boundaries (cities), at 15min intervals, identified by a hitherto unused highly accurate geolocation database, we are able to provide the first accurate estimate of the evolution of the internet's expansion at monthly intervals.

\begin{figure*}[p]
\captionsetup{width=11.4cm}
\begin{center}
\centerline{\includegraphics[width=11.4cm]{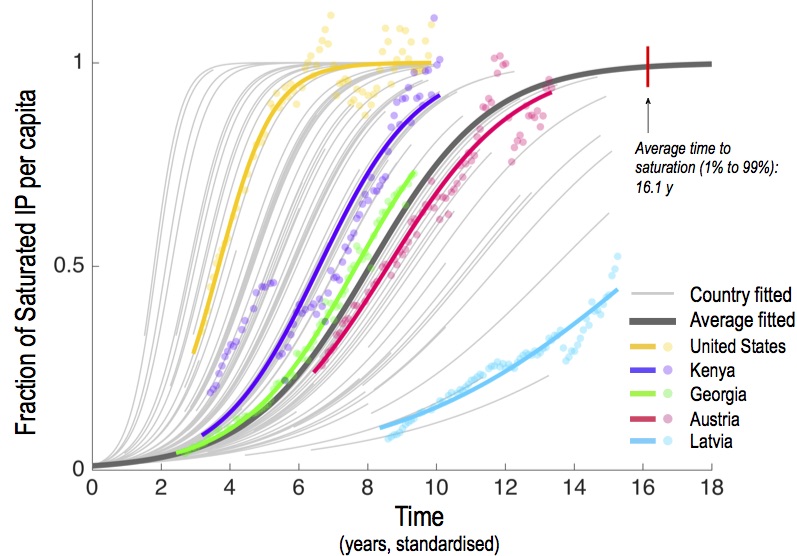}}
\caption{\textbf{The diffusion of the internet across 100 nations.} Thick \textit{main curve} shows the mean field logistic diffusion model fit to all 100 nations over an idealised 18 years of experience, beginning at 1\% saturation. Average saturation was estimated as 3 IPs per person. Red vertical bar shows the crossing point of the mean field line with 99\% saturation, occurring after 16.1 years of experience. Thin lines give normalised individual country fitted curves as variations from the mean field line, with example experience curves (coloured lines) shown together with the underlying, monthly, data (coloured markers). For all countries an x-offset was provided to correctly place the diffusion experience captured within the study’s 2006-2012 window of observation. Each curve has been standardised such that the initial fitted line intercepts the y-axis at 1\% saturation.}\label{figure2}
\end{center}
\end{figure*}

Significantly, given the temporal granularity and global scope of our series, we are able to confirm that the diffusion of the internet does indeed follow an S-, or logistic-, shaped process (Fig.~\ref{figure2}), mimicking studies of the diffusion of other technologies in the literature from hybrid corn\cite{Griliches:1957uq} to steam engines, electrification and personal computers\cite{Jovanovic:2005vv}. Accordingly, we estimate the temporal dynamics of IP per capita, $IPc$ at 1,647 cities globally as a logistic process $ IPc_{t}=K \big / 1+e^{-\alpha(t-\beta)}$, where $K$, $\alpha$ and $\beta$ are the asymptotic limit, the gradient and midpoint parameters respectively. We estimate this process as a non-linear mixed-effects model with a stochastic expectation maximisation algorithm (see {\bf S1}). By doing so, the algorithm is able to learn from the experiences of all countries by treating each country as a deviation (in time and gradient) from a generalised, or average, diffusion process.

We find that the internet's general diffusion process has an asymptotic limit of $0.32$ IPs per person, equating to an internet `saturation' level of approximately one IP address for a three person household, on average. Further we estimate that the diffusion process' average time to saturation within a country is just 16.1 years (1\%-99\%), eclipsing the estimated 100- and 60- year saturation times for the comparable GPTs of steam-power and electrification respectively\cite{Jovanovic:2005vv}. Our method also enables the elaboration of individual country experiences of the internet's penetration (see Table A, {\bf S1}). Our estimates reveal that whilst several nations already experience saturated internet penetration, others will not reach this point for decades.


\section{Sleep}

Next, we demonstrate the use of intra-diurnal variation in IP activity to create time-to-sleep, time-to-wake, and total overnight sleep duration estimates for 645 cities over a 7 year period. Recently, the internet's impact on waking human behaviour and on the duration and quality of sleep has come into focus\cite{Choi:2009dr,Lam:2014eb}, with laboratory evidence now confirming the impact of recent e-technologies on human chronobiology\cite{Chang:2015jr}. Unsurprisingly, prominent authors have been calling for a `broad data-collection strategy' to, `transform our understanding of sleep'\cite{Roenneberg:2013hc}. Whilst sleep scientists have seen the potential for the internet's use as a chronobiology data platform, they have so far conceived of this potential through traditional, self-reported time-use survey methodologies, albeit at scale\cite{Walch:2016fa}. 

We approach the problem differently. Our approach begins with the simple intuition that the switch from an internet-enabled device being offline to online at the beginning of the day, or conversely, online to offline at the end of the day, is correlated with the moment a person ends or begins their sleep. The association need not be exact, instead a systematically leading or lagging relationship carries the required information. Using this assumption, we first convert our geolocated IP activity dataset into intra-diurnal activity traces for each city. Next, we apply a novel machine-learning (ML) procedure to globally up-scale the highly detailed time-use survey data contained in the American Time Use Survey (ATUS) for 81 US cities having populations greater than 500,000 for all years which coincide with our internet activity dataset (see Section 2, {\bf S1}). By $n$-fold cross-validation, our method obtains an average error of 11~min. In effect, our approach converts granular internet activity data into a passive chronobiology monitoring platform at global scale. 

\begin{figure*}[p]
\captionsetup{width=11.4cm}
\begin{center}
\centerline{\includegraphics[width=11.4cm]{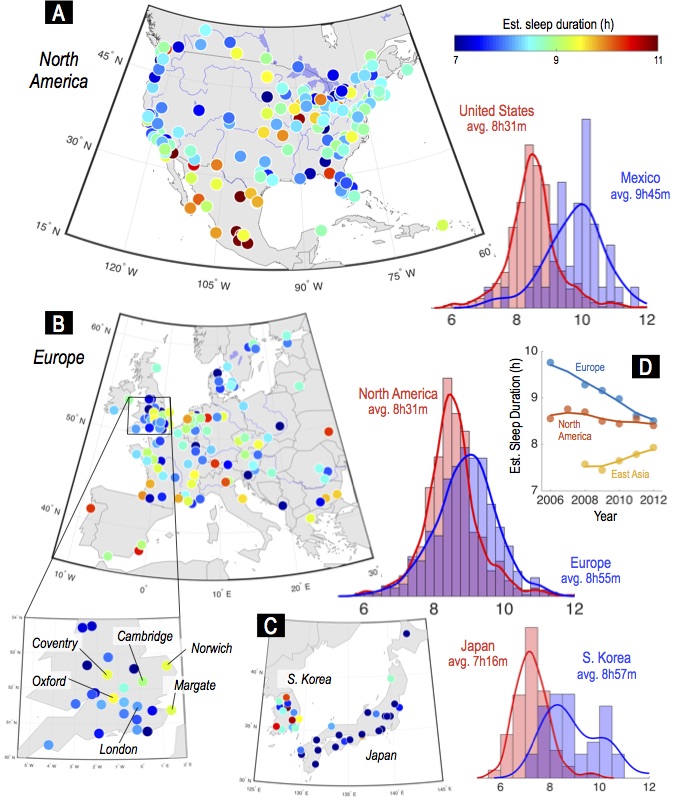}}
\caption{Estimating global sleep duration from internet activity. Machine learning was applied to average annual daily IP activity data for 81 US cities of population greater than 500,000 for which stratified American Time Use Survey time to sleep and time to wake data were available. The cross-validated model (mean error: 11m) was then applied to intraday IP activity data for 645 cities globally during 2006-2012 to produce estimates of sleep duration (the difference between times to wake and sleep). Estimated sleep duration heat maps during 2012 are shown for several regions (left panel). Density comparisons are based on all years (right panel).  \textbf{(a)} North America (left) with distributions plotted for all US and Mexican cities indicating that residents of US cities obtain 1h15m less sleep than their Mexican counterparts. \textbf{(b)} European cities (left) is shown together with southern England (detail). Compared to North America, European residents obtain approx. 25m more sleep, on average. \textbf{(c)} East Asian comparison, revealing Japanese obtain more than 1h40m less sleep than their South Korean neighbours. All two-way comparisons reported were significant at the $p < 1\times 10^{-10}$ level (two-tailed). \textbf{(d)} Changes in estimated sleep duration by UN regional classification indicating a potential convergence in sleep culture in developed regions globally.}\label{figure3}
\end{center}
\end{figure*}

We find that the sleep predictions differ statistical significantly across regions, hinting at an underlying cultural explanation (figure \ref{figure3}). In general, major cities tend to have longer sleeping times compared to surrounding satellite cities. Further, a comparison of estimated sleep duration aggregated at the UN regional scale over time suggests a sleep duration convergence phenomenon may be at play: whilst North America has remained largely static over the study window, Europe sleep duration has declined, and East Asian sleep duration has grown, reducing regional differences in sleep duration. Our approach uses only a single category in the ATUS dataset concerning sleep, however, over 100 tier 2 activity categories are available, emphasising the potential for IP-activity up-scaling across numerous research domains.


\section{Income and Productivity}
Finally, we show that IP activity data can be used to predict local economic activity as well as differences in sectoral productivity. This application reveals that, in a more aggregated form, highly granular IP activity data can be used to predict the outcomes of very complex human behavior and interactions. Our approach relates to a small but growing body of literature that uses other passively collected data to measure local economic activity \cite{Henderson:2012,Chen:2011,Jean:2016,Blumenstock:2015}, and a recent study which uses an estimate of aggregate IP allocation at the sub-national level to study digital ethnic favouritism\cite{Weidmann:2016jv}.

We use data from 411 large regions from middle and high income countries for the years 2006-2012. The regions are defined by the OECD and normally correspond to the first subnational level (i.e. U.S. states or EU NUTS2 regions). A simple comparison of economic activity and internet penetration between different regions is likely to be confounded by a number of other factors that drive economic and internet activity simultaneously (i.e. technological development, culture, geography etc.). Instead, we apply a fixed effects estimator that exploits the time-series features of our data and compares changes in economic and internet activity within the region over time.

Our measure for economic output is the regional Gross Domestic Product (GDP) per capita (in logs) in a given year, the measure for sectoral productivity is the Gross Value Added (GVA) per worker in a given year, and our measure for internet activity is regional IP per capita in a given year. In our estimation approach we account for time-invariant differences in economic development and productivity in a region, shocks that are common to all regions in a country and year as well as region specific linear trends.

\begin{figure*}[p]
\captionsetup{width=8.7cm}
\begin{center}
\centerline{\includegraphics[width=8.7cm]{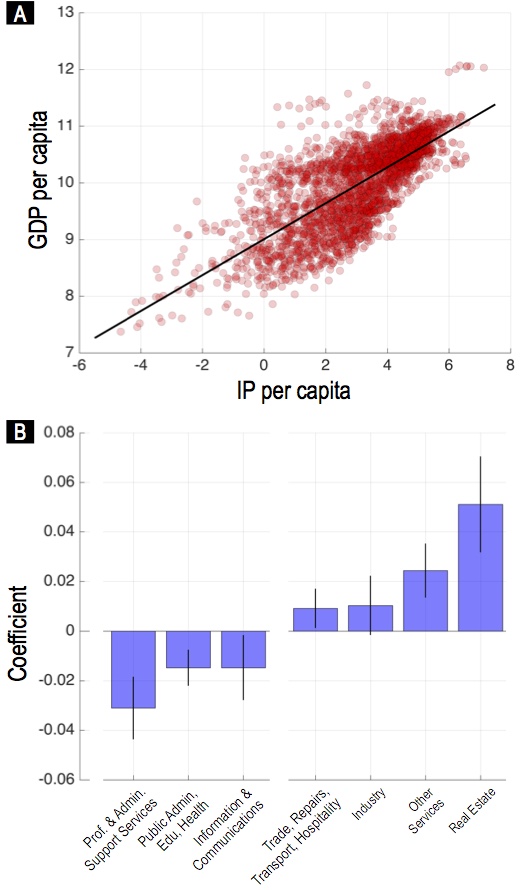}}
\caption{\textbf{Internet access and economic outcomes at the sub-national level.} \textbf{(a)} GDP per capita versus IP addresses per capita accounting only for year fixed-effects in $411$ sub-national regions defined by the OECD (i.e. states in the USA or NUTS2 in Europe) over the years 2006-2012  ($n$ = $2,832$; linear trend, $R^2$ = $0.61$). \textbf{(b)} Mean estimated coefficients of the relationship between IP addresses per capita and each economic sector (bars). Error bars show estimates of the 95\% lower and upper bound confidence interval for the coefficient value across the 7 ISIC Rev.4 sectors (range of $n$ per sector, $1,041-1,422$, total $n = 7979$).}
 \label{figure4}
 \end{center}
\end{figure*}

We find a positive correlation between GDP pc  and IP pc (Figure~\ref{figure4}(a)).  The simple correlation coefficient without accounting for region-specific, country-year specific differences and region-specific time trends is 0.38 (see Table E, {\bf S1}). Once we include those other covariates, the coefficient decreases to 0.08, suggesting that a 10\% increase in IP pc is associated with a 0.8 \% increase in GDP pc at the regional level. However, as Figure~\ref{figure4}(b) makes clear, increased internet activity is not associated with uniformly positive impacts on all economic sectors within a region. Broadly speaking, we find that service sectors amenable to digital competition through out-sourcing (e.g. publishing, news, film production, administrative support, education) have suffered with increasing local IP concentration, whilst location-constrained sectors have prospered from higher internet concentrations presumably due to lowered consumer search-costs and/or logistic and process efficiency gains (e.g. wholesale, retail, real-estate, repairs, hairdressing, mining, transportation, accommodation) (see Table F, {\bf S1}). It is important that the estimated effects on regional GDP and sectoral GVA, respectively, are only correlations and  do not allow for a causal interpretation.

\section{Conclusion}
To our knowledge, the present study is the first of its kind to apply over a trillion online/offline activity observations of the entire internet to human behaviour. The data's high level of spatial and temporal granularity paired with the passive way it is collected, makes IP data uniquely suited to analyse a wide spectrum of human behaviour and social interactions. As such, our work not only expands the data and methodological space of the quantitative social data-sciences but it provides a first glimpse of the potential of global internet activity to change profoundly the way research in this realm is conducted.


\newpage 

\paragraph{Acknowledgments}
We thank John Heidemann and the group at the USC, PREDICT program who provided the IP Activity data and technical support. We thank Shlomo (Solly) Angel at NYU-Stern for providing population data from the Atlas of Urbanization.  This work was in part supported by auDA Foundation grant, `A new, high spatial-resolution, dataset on internet use in Australia' (2013).

\paragraph{Author Contributions} All authors conceived and designed the study.  
KA conducted the major cluster join and spatial aggregation procedures. All 
authors analysed the data, prepared figures and wrote the appendix 
information.  SA and PR wrote the paper. All authors edited and commented on 
the paper.


\newpage
\appendix

\section{Related `big data' social science studies}
 A small, but rapidly growing number of studies has demonstrated the utility of passively collected `big data' for the pursuit of social data science. Like the present work, these studies leverage the \emph{passive} nature of the data -- be it the analysis of cell-phone meta-data~\cite{Onnela:2007ek,Blumenstock:2015,Toole:2012jh}, `app' activity logs~\cite{Ginsberg:2009ep,Franca:2015dv,Bakshy:2015jn}, or night-time satellite imagery~\cite{Jean:2016,Chen:2011,Henderson:2012}. In each of these studies, the data were not specifically collected for social data-science purposes, instead the data arose as a by-product of the particular service delivery in question and has been creatively applied to long-standing (and at times, entirely novel) social science research questions. Together, these papers demonstrate the variety and richness of scientific insights now available in a world of `big-data'.
 
 Of these contributions which concern cell-phone meta-data, such as is used in~\cite{Onnela:2007ek,Blumenstock:2015,Toole:2012jh} the underlying data are presently only available through a dedicated agreement with a commercial cell-phone provider. Such sources are typically spatially restricted owing to the regional focus of cell-phone companies. Similarly, datasets of this nature are also time-bound as the companies involved prefer to release only a limited temporal segment of their data at a time.
 
  Likewise, software-based (`app') data of the kind reported in~\cite{Ginsberg:2009ep,Franca:2015dv,Bakshy:2015jn} which concern respectively Google search, Twitter, and Facebook, present fascinating opportunities for the study of novel real-time health monitoring, time-use, and political polarisation. Given the high engagement of interent users with some of these applications, it is possible that through open-data policies, one might be able to replicate the kind of work we report here. However, there are inherent compromises with such data beyond the commercial nature of the source. For instance, these data are subject to sample selection bias of at least two kinds. First, it is unlikely, even for the most popular `app', that a one-to-one mapping will exist between the app's user base and the universe of the internet's user base.  Second, since apps are always crafted for a particular use-case, an app's activity data will necessarily be restricted to a user's engagement profile within the app in question.
  
  Satellite imagery presents a third approach to big-data in the social sciences~\cite{Jean:2016,Chen:2011,Henderson:2012,Raschky:2014vm}. Here, the data are inherently global in scope, but limited in temporal granularity with the typical observation being an annual reading of  the night-time luminosity of 1 square kilometre of the Earth's surface.
  
  Finally, we mention the recent work of \cite{BenitezBaleato:2015hk, Weidmann:2016jv} who leverage internet `activity' to gain interesting insights on apparent ethnic bias in the allocation of IPs at the subnational level. Here, like cell-phone studies mentioned above, the data arise from a partnership with a commercial Internet Service Provider (ISP) and so exhibit similar user-, and temporal-, restrictions. In this case, the ISP provided two, 16 day, contiguous segments of its customer internet activity, for each of the years during 2004 to 2010. Data were spatially aggregated to ethnographic regions (GeoEPR~\cite{Wucherpfennig:2011gu}) by using the MaxMind GeoIP2 City IP-location database, which, by the provider's own estimates contains some significant errors\footnote{See \url{http://maxmind.com/en/geoip2-city-database-accuracy}. For instance, MaxMind report that just 14\% of IP locations to cities in Australia are `correctly resolved' within their dataset.}, especially in the developing world. Nevertheless, uniquely active IP blocks\footnote{A block in this work is at the $/24$ level, meaning identification down to the second-to-last integer of a standard IPv4 IP address. For example, 192.172.3.xxx .} are identified and used to assess the level of bias in their spatial allocation.

 Thus, we view node-to-node online/offline scan data of the kind used in the present work as complimentary to these other passive data sources for the progress of quantitative social data science. Whilst online/offline activity traces are not subject to the same kinds of user- or temporal- sample selection bias as mentioned above, the dimensionality of the data gained through IP scanning is a fraction of what might be available in a single `observation' available in cell-phone call meta-data or an app's session history. Nevertheless, as demonstrated by the breadth of applications reported in this work, low dimensionality does not prevent the leverage of the unprecedented scope and granularity of IP scan data for scientific inquiry.
 
 Moreover, it is worth noting that online/offline scan data is unique in one, further, and particularly striking characteristic: with sufficient know-how, the data may be gathered by \emph{anyone} who is online -- no government agency, nor private company, nor intermediary of any kind is required to enable the collection of IP online/offline data\footnote{For example, consider the internet survey conducted by Carna Botnet, \url{http://internetcensus2012.bitbucket.org/paper.html}.}. For this reason, for so long as the internet retains its' democratic foundations, we consider online/offline data of the kind employed here as fundamentally `open'.

\subsection{Data sources}

Internet activity data were provided by the University of Southern California (USC) PREDICT internet-security database whilst IP-geolocation information was provided by a highly accurate commercial source. Specifically, we utilised USC PREDICT's IP activity full census of all $2^{32}$ IP addresses as well as 1\% sub-sample scans which provide repeated online/offline observations for clusters of IPs at 11~min intervals\cite{Heidemann:2008cv}. Here, the most basic node-to-node query (a `ping') is sent, asking the target IP if it is presently online, returning a success indicator and return time.

In the case that an IP address is not online, or unreachable due to firewalls or other prohibitions, the nearest router or host will respond. Our method aggregates these scans to 75 million rows of online/offline IPs ($\#^{\text{on}}$, $\#^\text{off}$), for 15min intervals, at 1,647 urban-boundaries in 122 countries\cite{angel2010planet}. Significantly, our data cover a key phase of the internet's global expansion with the user base doubling from roughly 16\% to over 35\% during 2006-2012\cite{ITU:2016tu}.

Further spatial and temporal aggregation methods are discussed below relative to the specific scientific application.

\section{Part 1: Measuring the diffusion of the Internet}\label{sc:diff}
\begin{figure}[tbp]
\begin{center}
         \includegraphics[angle=0,width=0.9\textwidth]{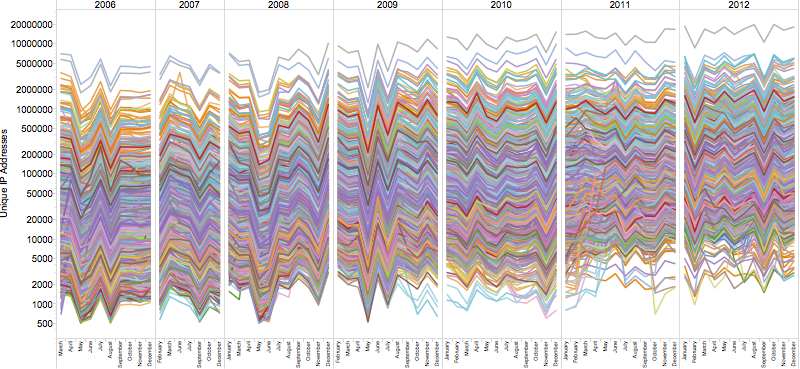}
\caption{IP space utilization by month and city}
         \label{ip_raw}

\end{center}
\end{figure}

The data used for this section was created following the same procedure as is described in the data descriptor, albeit with a difference in the temporal identification window. Instead of 15-minute intervals, counts of unique IP addresses assigned to a location were identified over a month of observations. A month was used as the time window to capture the dynamics of the IP space, as well as account for the differences in scanning frequency during the period. We experimented with different geographical and time window aggregation until a consistent pattern emerged. Figure \ref{ip_raw} presents the resultant monthly IP space utilisation (count of unique IP addresses) per city over 2006-2012.  For ease of comparison, line colouring is grouped for cities within the same country. The emerging pattern has similarities to a \textit{missoni} colour scheme used in fashion. Months with no scanning activity or less proportional scanning compared to the other months are excluded.

\subsection{Correction for missoni bias}

\begin{figure}[tbp]
\begin{center}
         \includegraphics[angle=0,scale=0.5]{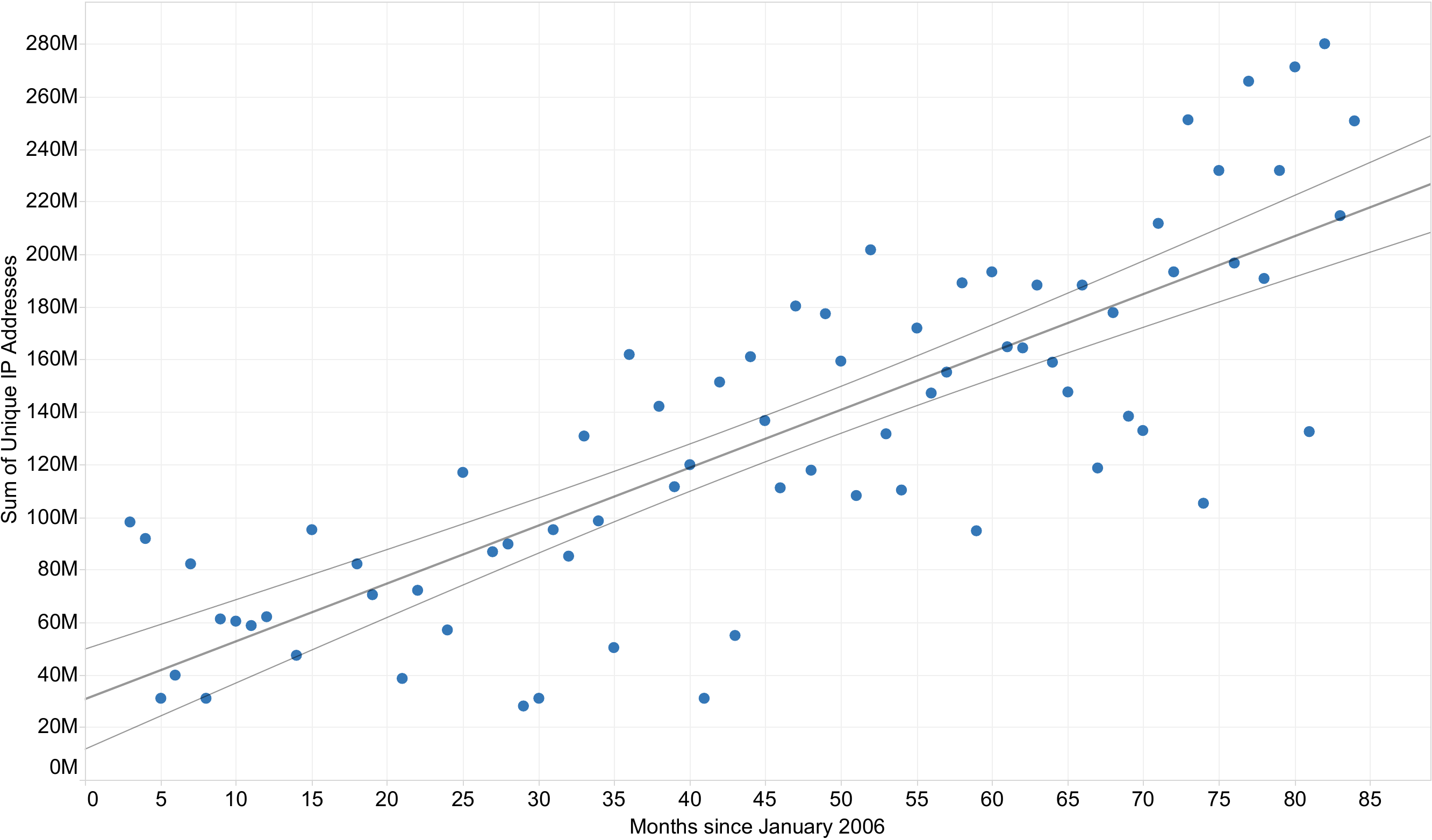}
\caption{Sum of all unique IP addresses by month}
         \label{sum_ip}

\end{center}
\end{figure}

\begin{figure}[tbp]
\begin{center}
         \includegraphics[angle=0,scale=0.5]{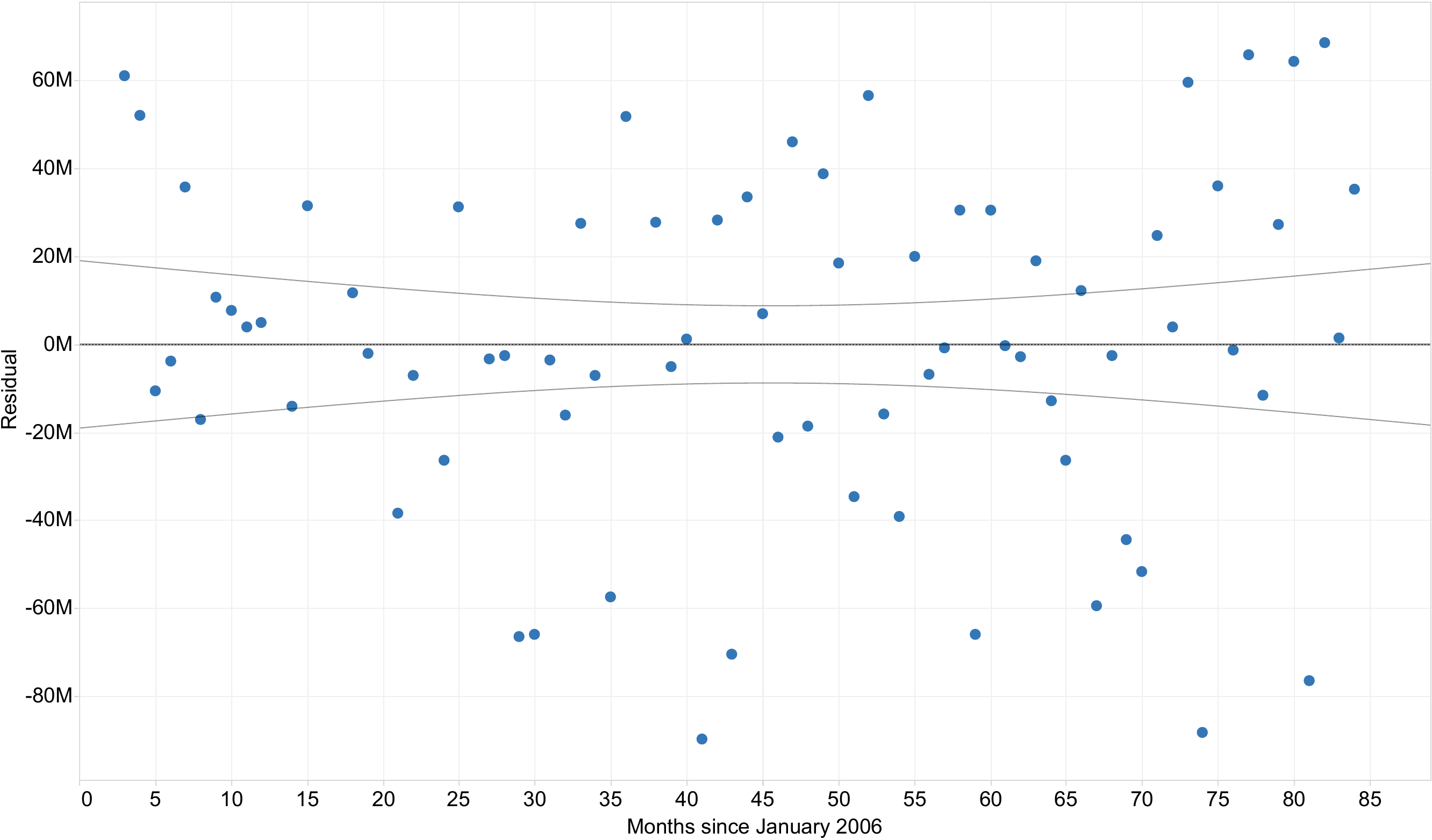}
\caption{Residual plot of the linear trend fitted for the correction of the missoni bias}
         \label{residual}

\end{center}
\end{figure}

\begin{figure}[tbp]
\begin{center}
         \includegraphics[angle=0,width=0.9\textwidth]{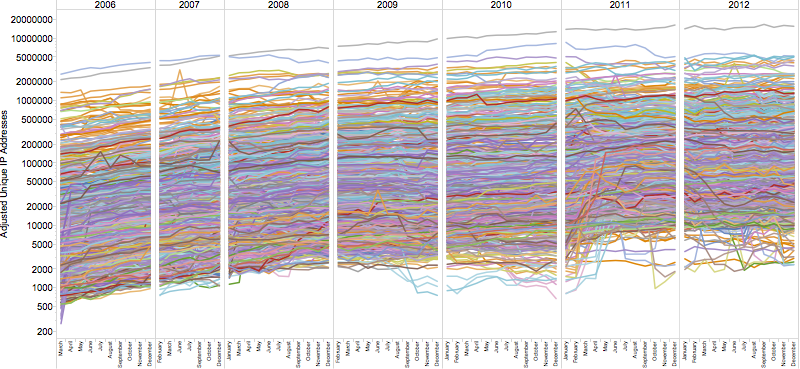}
\caption{Adjusted IP space by city and month}
         \label{ip_adj}

\end{center}
\end{figure}

To consistently estimate the growth of the internet over time corrections must be made for the random timing of the IP scans along with the general dynamics in the IP space. First, we filter out all cities which do not have consistent measurements across all months. Visually we identified a minimum cut-off 500 IP addresses per month, leaving $829$ cities. This threshold served to discard individual city IP traces which followed an erratic pattern not consistent with the trend of the majority. Second, we sum up all unique IP addresses over all remaining cities per month and fit a linear trend as shown in figure \ref{sum_ip}. The residuals of this regression are shown in figure \ref{residual}. The residuals show no sign of pattern or trend that would be removed from the data when assuming a linear trend. Third, for each month we calculate a generic scaling factor, which would move individual monthly values onto the linear trend line. Finally, we apply the generic monthly scaling factor to each city month observation, and so, remove the missoni bias. Results of this procedure are given in figure \ref{ip_adj}.

 After processing, 576 cities were retained which had consistent measures of IP activity and population data from which monthly IP {\it per capita}  measures were constructed. Finally, in this report, we aggregate at the country level by averaging across cities. Together, we have 75 monthly average IP {\it per capita}, $IPc_t$  observations (Jan 2006 to Dec 2012 with excluded under-sampled months) for each of 122 countries, comprised of between 1 (various) and 70 (e.g. USA) cities.

\subsection{Estimation of the Urban Internet diffusion process}
We use a standard logistic function for estimating the diffusion process. Diffusion processes in Economics are usually expressed as adoption rates in percentages as in the case for hybrid corn \cite{Griliches:1957ks}. We use population\footnote{We are very grateful to Prof. Solly Angel and his team at the NYU Stern Urbanization Project for sharing with us the pre-released population data of their new Atlas of Urbanization. } by city to derive a comparable measure of IP per capita with each defined urban boundary. Population data are available for 2000 and 2010 for future revised city boundaries. We use the population data for all cities, where the city definition did not change. Cities that previously were separate and later defined as a unified urban zones or cities which were split are excluded. 576 cities worldwide remain with consistent measures of IP and population data. We interpolate and extrapolate the demographic data linearly between 2006 and 2012.

More developed countries have a higher proportion of cities in the sample, such as the United States (70) or the Russian Federation (42), while other countries only have one city like Zimbabwe or Vietnam. We average the IP per capita values by month by country to derive a unified measure of urban IP per capita. The model estimated is the logistic growth curve
\begin{equation}\label{P.2_eq.1}
IPc_{t}=\frac{K}{1+e^{-\alpha(t-\beta)}}
\end{equation}
which was estimated as a non-linear mixed-effects model with a stochastic expectation maximisation algorithm. $IPc_{t}$ is IP per capita in a given month, $K$ the asymptotic limit, $\alpha$\footnote{In the estimation using Matlab function \textit{nlmefitsa} we use the inverse of $\alpha$ as parameter transform to have similar scales as $\beta$ to help the simulation to converge.} the gradient and $\beta$ the midpoint. Assuming that the global diffusion process of the Internet takes place with each country contributing different stages of the full curve and by assuming that each country contributes to a general diffusion process, we can estimate an average curve and country specific trajectories. We estimate the listed parameters of the general process, as well as allow for cross dependent random offsets for each country. We used many simulations to achieve an RMSE of $0.0143$. 

\subsection{Results and Country rankings }
\begin{figure}[tbp]
\begin{center}
         \includegraphics[angle=0,scale=0.25]{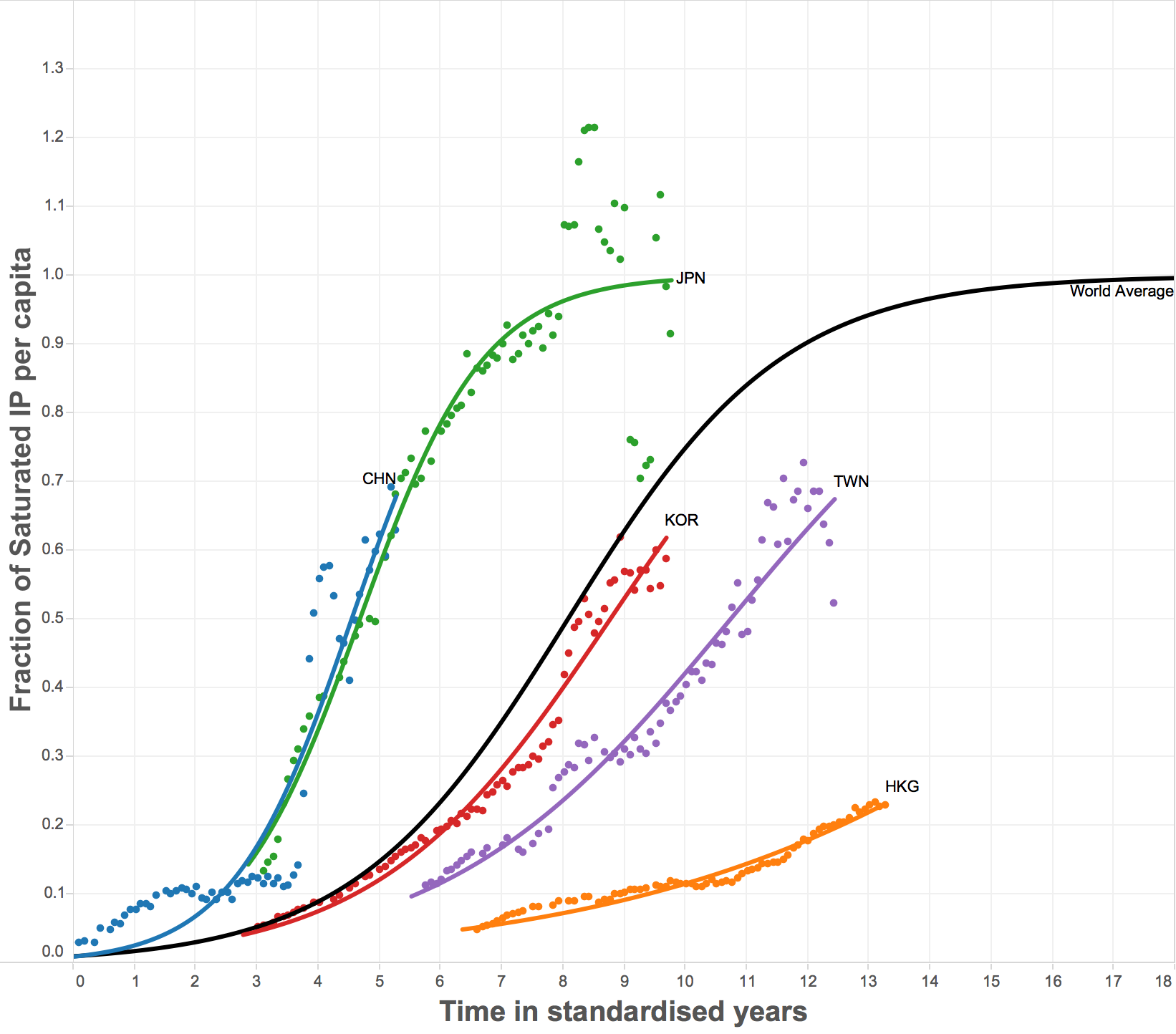}
\caption{The diffusion of the internet in Eastern Asia}
         \label{ipdiff_eastern_asia}

\end{center}
\end{figure}

\begin{figure}[tbp]
\begin{center}
         \includegraphics[angle=0,scale=0.25]{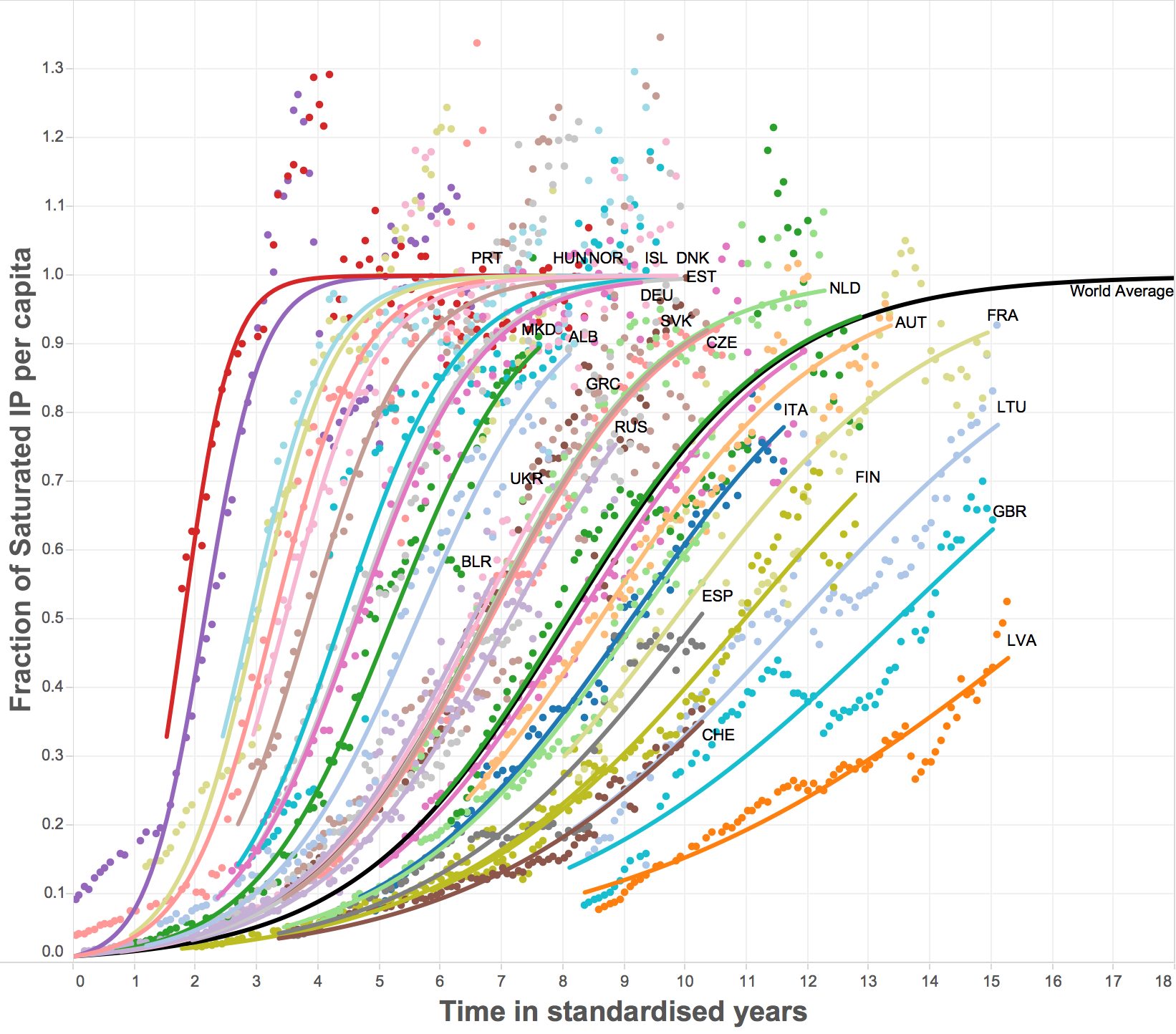}
\caption{The diffusion of the internet in Europe}
         \label{ipdiff_europe}

\end{center}
\end{figure}

\begin{figure}[tbp]
\begin{center}
         \includegraphics[angle=0,scale=0.25]{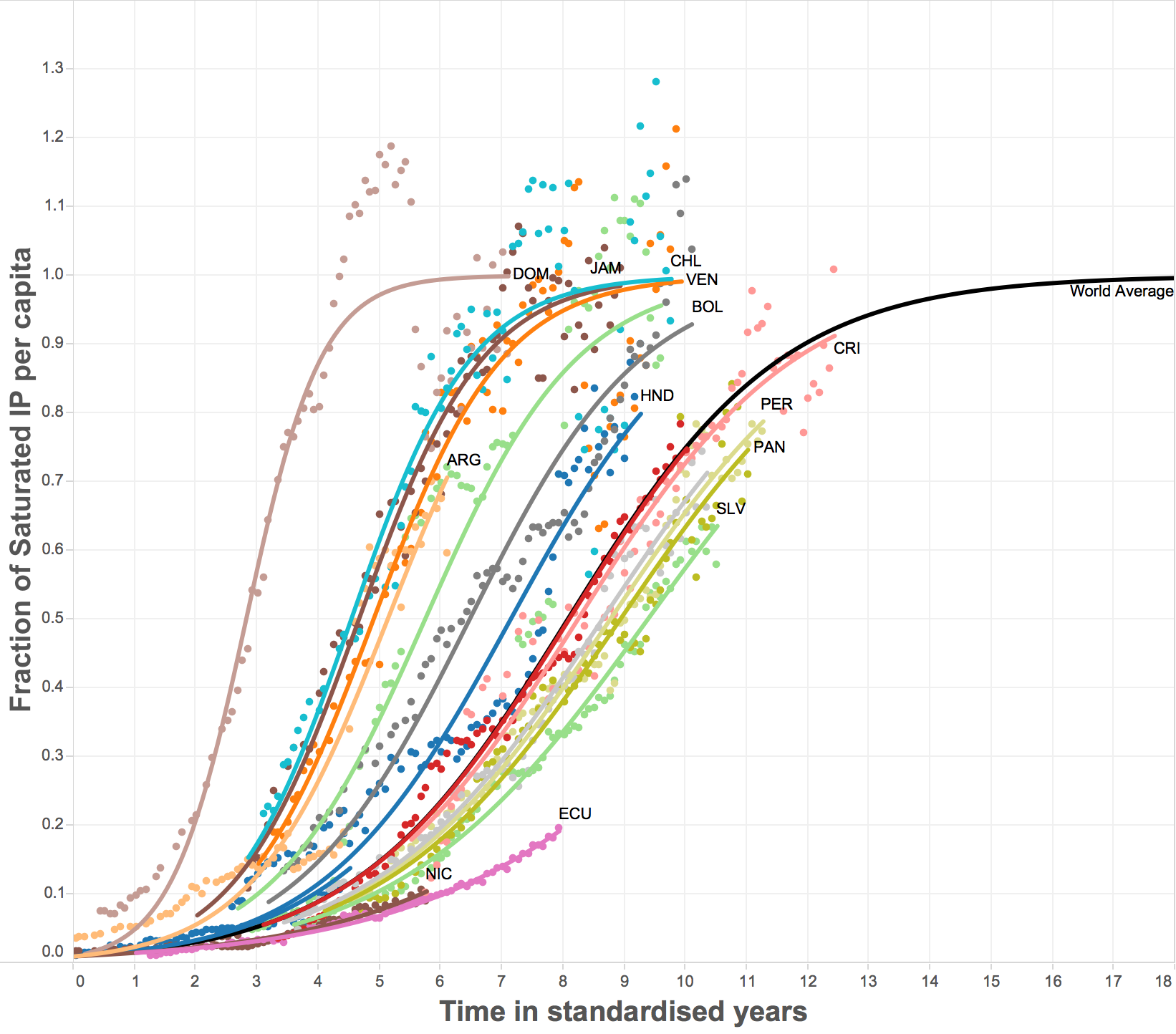}
\caption{The diffusion of the internet in Latin America}
         \label{ipdiff_latin_america}

\end{center}
\end{figure}

\begin{figure}[tbp]
\begin{center}
         \includegraphics[angle=0,scale=0.25]{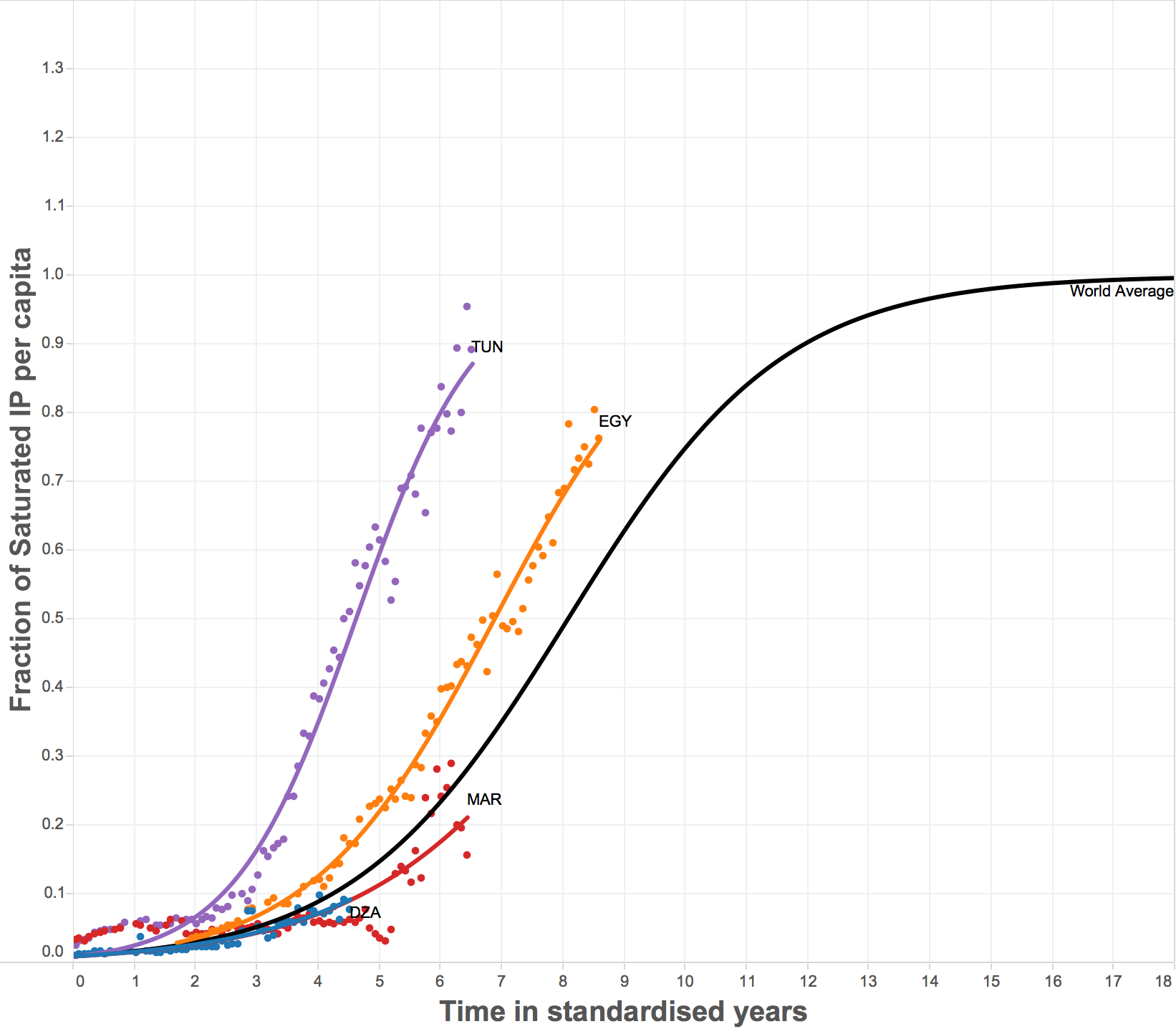}
\caption{The diffusion of the internet in Northern Africa}
         \label{ipdiff_northern_africa}

\end{center}
\end{figure}

\begin{figure}[tbp]
\begin{center}
         \includegraphics[angle=0,scale=0.25]{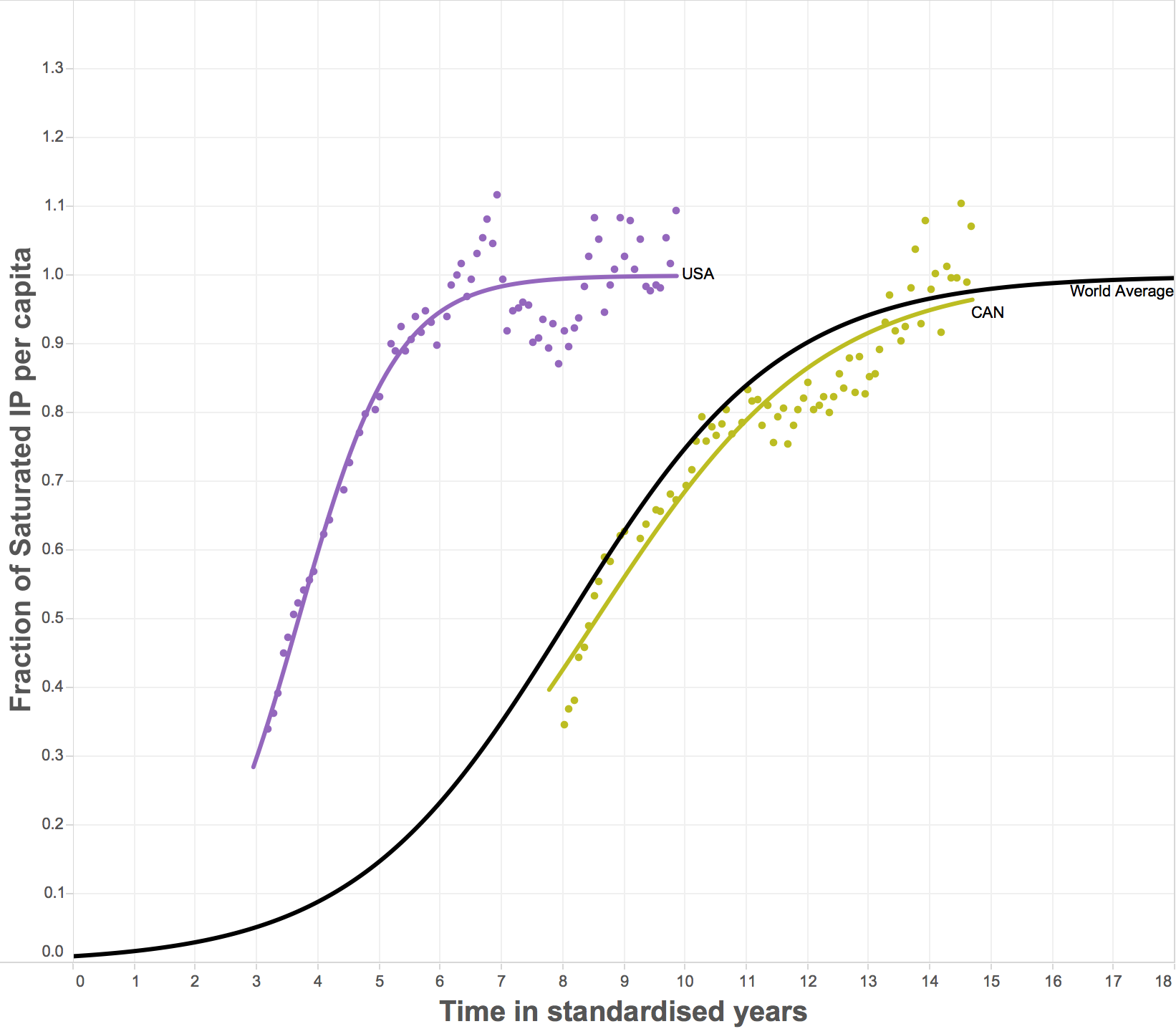}
\caption{The diffusion of the internet in Northern America}
         \label{ipdiff_northern_america}

\end{center}
\end{figure}

\begin{figure}[tbp]
\begin{center}
         \includegraphics[angle=0,scale=0.25]{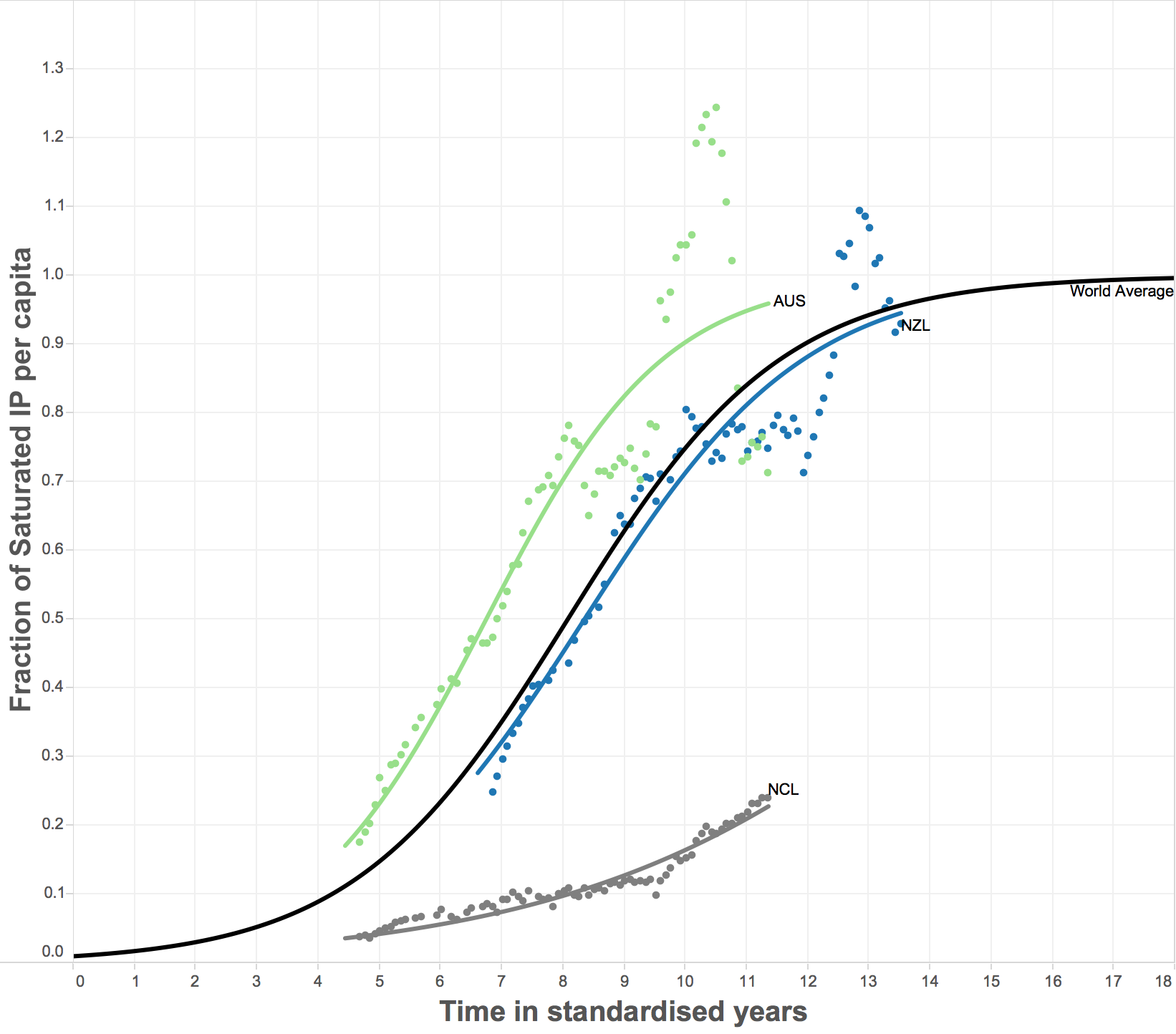}
\caption{The diffusion of the internet in Oceania}
         \label{ipdiff_oceania}

\end{center}
\end{figure}

\begin{figure}[tbp]
\begin{center}
         \includegraphics[angle=0,scale=0.25]{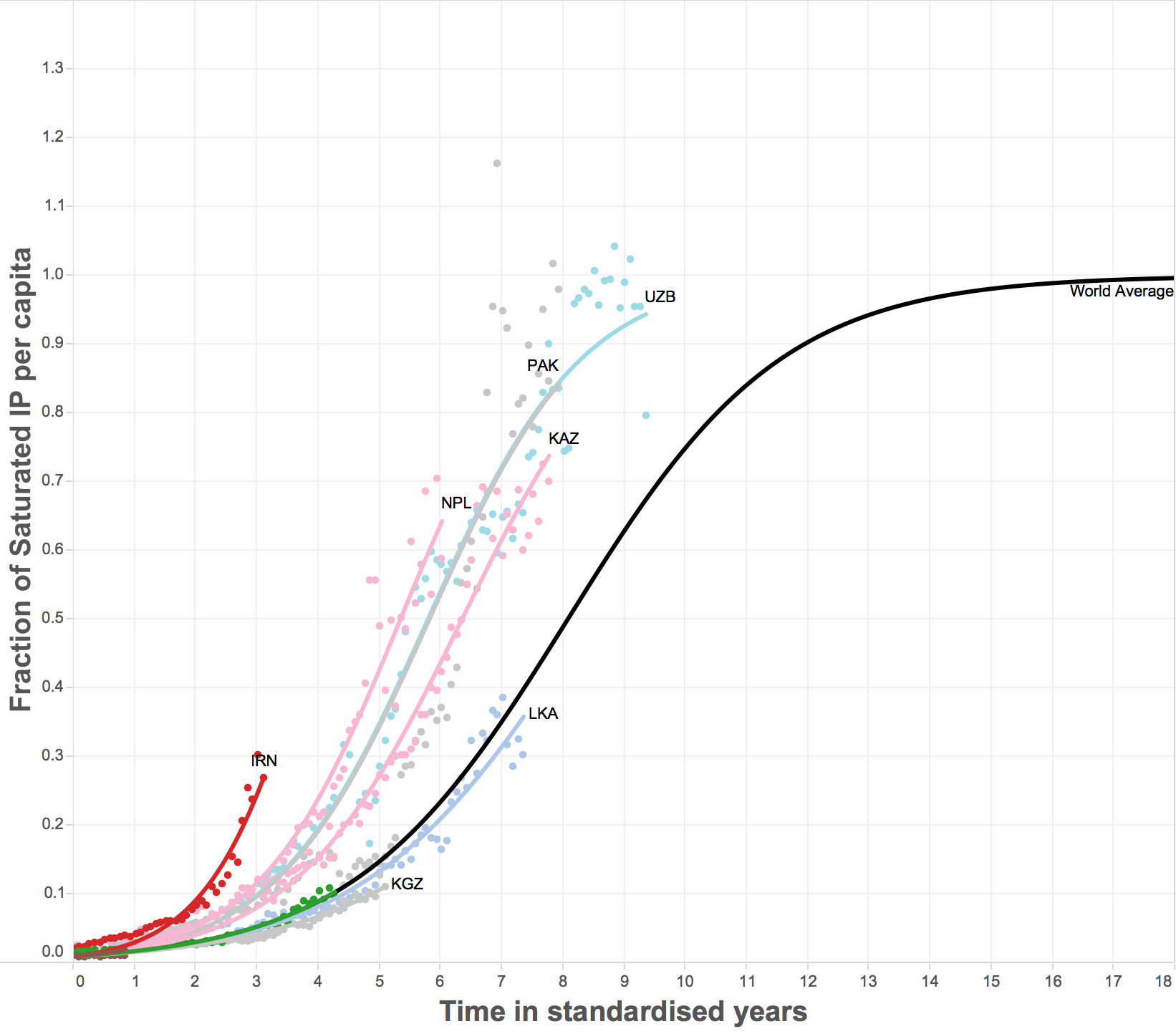}
\caption{The diffusion of the internet in South Central Asia}
         \label{ipdiff_south_central_asia}

\end{center}
\end{figure}

\begin{figure}[tbp]
\begin{center}
         \includegraphics[angle=0,scale=0.25]{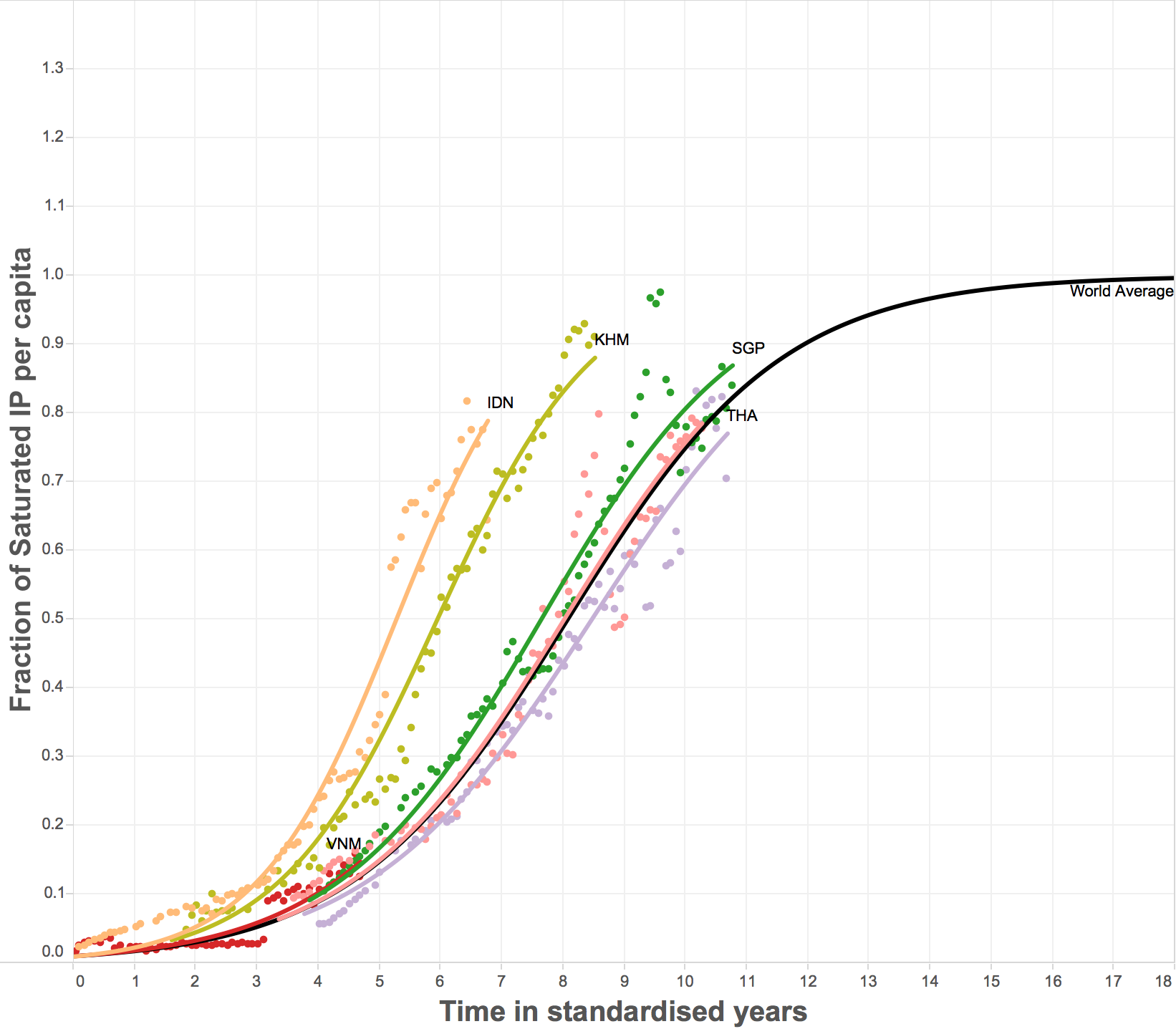}
\caption{The diffusion of the internet in South Eastern Asia}
         \label{ipdiff_south_eastern_asia}

\end{center}
\end{figure}

\begin{figure}[tbp]
\begin{center}
         \includegraphics[angle=0,scale=0.25]{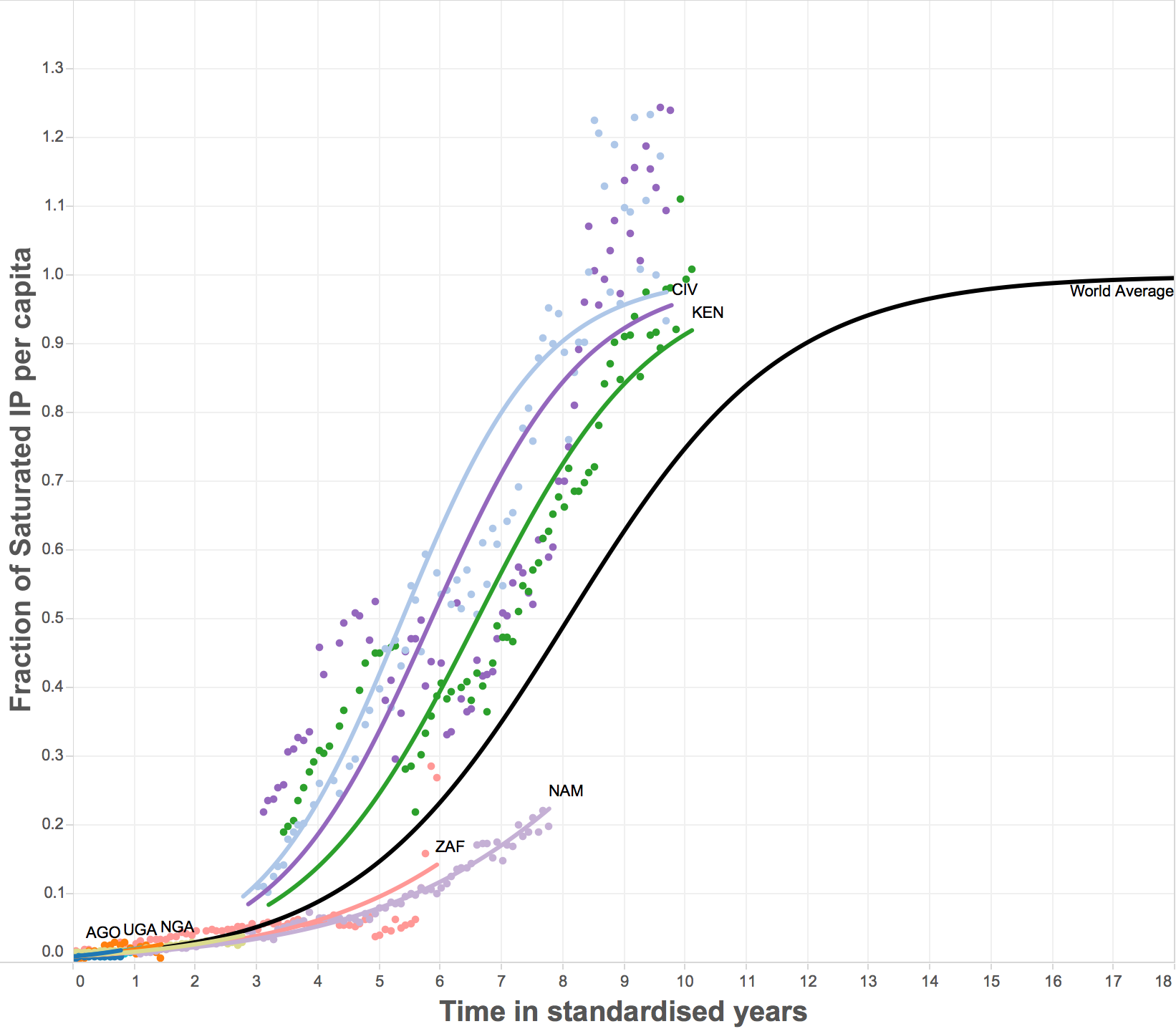}
\caption{The diffusion of the internet in Sub Saharan Africa}
         \label{ipdiff_sub_saharan_africa}

\end{center}
\end{figure}

\begin{figure}[tbp]
\begin{center}
         \includegraphics[angle=0,scale=0.25]{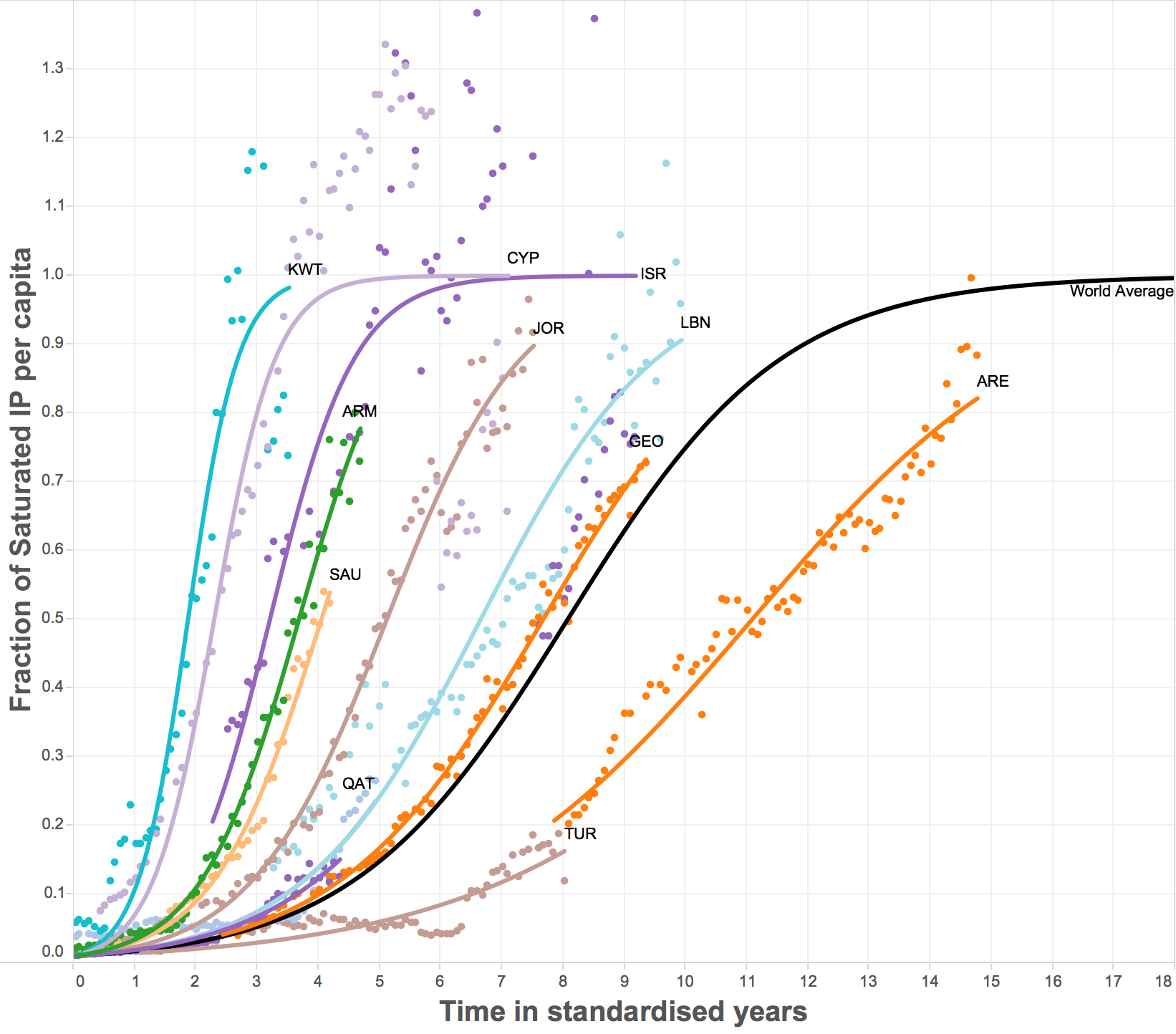}
\caption{The diffusion of the internet in Western Asia}
         \label{ipdiff_western_asia}

\end{center}
\end{figure}

The general asymptotic limit is estimated to be $0.32$ which approximately represents one IP address per household unit with three people on average. We defined the diffusion duration as the range from 1\% to 99\% since even low levels of Internet penetration, such as might be present in public libraries in a given city, already represent an information shock. The estimated average diffusion of the Internet in metropolitan areas is estimated to be $16.1$ years, with a growth rate of $0.047$.  We categorized the IP per capita level starting with the category \textit{Saturated} which corresponds to an IP per capita value in 2012 above the overall average of $0.32$ IP per capita and then creating subsequent category thresholds by successively halving this level (\textit{High,Medium,Low}). Table \ref{ip-cap-tab} ranks all countries based on their urban IP penetration in 2012 compared to their levels in 2006. For each country we present the estimated asymptotic limit, growth rate and corresponding years equivalent to either 1\% or 99\% of saturation for their particular growth dynamics. Figure \ref{ipdiff_eastern_asia},\ref{ipdiff_europe},\ref{ipdiff_latin_america},\ref{ipdiff_northern_africa},\ref{ipdiff_northern_america},\ref{ipdiff_oceania},\ref{ipdiff_south_central_asia},\ref{ipdiff_south_eastern_asia},\ref{ipdiff_sub_saharan_africa} and \ref{ipdiff_western_asia} plot the same observation as figure 2 in the main article split across united nation regions.

The country rankings represent different technological development outcomes as well as consequences of large scale government investments or their absentia. The utilisation, or consumption in Economic terms, of the IP space is dependent on the underlying physical technology used to connect to the Internet. During the period from 2006 and 2012 the main technological approach to internet connections was via fixed line connection. According to the raw unfiltered location data, mobile IP addresses make up 0.1\% at the end 2008 and 5\% at the end of 2012. In the top \textit{Saturated} group Germany is leading the list but in contrast to South Korea, the last in that group, Germany is expected to reach is asymptotic limit in 2013, while South Korea is on a trajectory of IP per capita growth until 2021. Notably, Macedonia, ranked 8th in 2012 after climbing 43 places since 2006, is likely showing the outcome of a massive investment in Internet infrastructure through a USAID program (`Macedonia Connects'\footnote{See \url{http://bit.ly/1uieJb9}.}) initiated in 2007~\cite{hosman2010policy}. Likewise Estonia, ranked 3rd, has a widely-known digital story, being a vigorous adopter of ICT with high-profile software products from Estonian developers such as Skype\footnote{See \url{http://bit.ly/2cjJcTH}} being a most visible by-product. In last place on the list is Angola followed by Cote d'Ivoire, though Angola seems to be on a path to IP per capita growth in contrast to Cote d'Ivoire. 
  
\begin{center}
\begin{scriptsize}
\input{ip_cap_table.tex}

\end{scriptsize}
\end{center}

\newpage

\section{Part 2: Estimating global sleep patterns by ATUS up-scaling}\label{sc:wsc}

\subsection{Data}
The data used for the prediction of sleep is taken from the American Time Use Survey (ATUS). The ATUS is a survey conducted on a sub-sample of the Current Population Survey (CPS). Specific respondents are picked after being surveyed for eight months to have a representative time use allocation by region and demographics. The respondents are interviewed over the phone between 2 to 5 months after CPS has ended, to detail all their activities they conducted on the day before with start and stop times. A prominent activity in the survey is a respondent's sleep activity (sleep time and wake time).

We use \textit{Sleeping} category (ATUS: 010101) as our main input. Note, that the survey provides instructions for coding \textit{Sleeplessness} (ATUS: 010102), which we do not consider. The ATUS and CPS data are extracted using an online extraction builder that allows pre-selection of variables \cite{hofferth2013american}. The ATUS and CPS data are merged, as the metropolitan area \textit{fips} code is not readily available in the ATUS data files. For each respondent, we count the first time diary entry after 7 pm as sleep start and the last entry before 12 pm as sleep stop. Every response outside of this time frame was discarded (1,270 out of 218,609 cases). The average sleep start and stop time by metropolitan area is calculated by year using successive difference replicate (SDR) weights provided by the survey. We focus only on metropolitan areas with population exceeding 500,000 people in a given year as the census office warns that estimates with fewer people should be treated with care. Finally, this leaves 81 metropolitan areas in the US, which can be used as input for the wake-sleep cycle model.

For the internet activity data, we begin by calculating the IP \emph{fraction online}, $f_{i,t}$ for a given city, $i$ during a 15~min interval, $t$ as, $ f_{i,t} = {\#^{\text{on}}_{i,t}}\big / ({\#^{\text{on}}_{i,t}+\#^{\text{off}}_{i,t}}) $. Given that many servers are always online, and that workplaces typically connect to the internet with a single forward-facing internet address, the majority of variation in $f_{i,t}$ with which we are interested is generated by personal use. For the period in question, only a fraction of IPs in our dataset are assigned to cellular devices ($\sim$ 0.1\% in 2008, and $\sim$ 5\% in 2012), implying that intra-diurnal IP activity variation in $f_{i,t}$ will largely be due to personal internet activity.
%
%

To ensure adequate representation only city--years where at least 30 days per year with above 100 online IP addresses per 15min segment across each of the 96 contiguous segments of a day were retained leaving 645 cities in over 100 countries, or 1,648 city--years.

\subsection{The machine learning problem}

The aim of the wake-sleep cycle prediction is to classify internet traces automatically worldwide into predominantly awake or sleep periods by up-scaling a traditional time use survey. The average sleep start and stop time by city represents a continuous prediction problem. To overcome the limitations of having only 81 metropolitan areas and six years as observations for the forecast, we transform the problem. Instead of trying to predict the continuous time of an average person waking up or falling asleep, we convert the average values into the 15-minute time-slices consistent with internet data resolution. The outcome variable is set to \textit{1} for a segment (96 in total per day) when the segment is within the average sleeping time and \textit{0} otherwise. The problem becomes therefore, a classification problem: each city--year being represented by 96 binary rows.

\subsection{Feature generation}
To account for always-online servers and equivalently focus our inquiry on the \emph{variation} in fraction-online behaviour in each city--year, each daily trace (96, 15min segments) was first normalised to the [0,1] interval. Next, for each city--year, synthetic weeks were prepared by collecting all days corresponding to each day of the week (e.g. all `Mondays', `Tuesdays', and so on). The average of all such days was taken to generate a representative week of seven days in sequence (`Mon', `Tue', ..., `Sun'). Finally, a robust smoother \cite{Garcia:2010hn} was applied to each synthetic week-day to account for any residual noise. A parameter setting of 500 for the smoorther was used (a strong setting for this particular procedure).  Together, these steps provided our first set of internet activity features: seven fraction-online columns, one for each day of the synthetic weeks of a given city--year combination.

Based on the synthetic fraction online, additional segment features were created. To capture the rates of change between the segments we include the first and second differences, producing a further 14 features. Consequently, we lose the first two 15min segments (after midnight) of each synthetic daily trace. Furthermore, we create dummy variables by weekday, which are 1 at the peak and trough of a day respectively, 0 otherwise (14 more features). As there might be different scanning behaviours and sleep respondent measures by year, we include a dummy for each year (6 features). In case important information is held within the full weekly synthetic trace, each synthetic week by city--year was compressed using wavelets (\textit{sym3}, level 7) providing 10 coefficients as further features. Finally, we include the absolute latitude of each metropolitan area as input (1 feature). Together, we have a total of 52 features from which the model can learn.

\subsection{Training and Testing}
$n$-fold cross-validation by city was used for training a machine learning algorithm. We used bagged (bootstrap aggregation) decision trees (``Random Forest'') for our prediction. By leaving out a city across all years for each training iteration, we train the model on the marginal city, which is the use case in mind when we apply the trained model to predict wake-sleep cycles worldwide. We achieve a classification accuracy of $0.9807$, using $0.5$ as prediction score threshold when we utilise all described features. In comparison, when we leave out latitude as input and just focus on Internet input data we get an accuracy of $0.9802$. We explored the tuning parameters of the algorithm, such as depth or leaf size, but we achieved the highest accuracy with feature engineering alone and relying on the standard random forest algorithm \cite{Breiman:2001fb}. The most important feature according to the cross-validation is the feature that represents the first difference of the fraction online on Monday followed by the feature with fraction online on Saturday as shown in table \ref{feature-importance}. The ATUS survey is a weighted average of activities on workdays and weekends, which is highlighted as an outcome of the feature ranking.

\begin{center}
\begin{scriptsize}
\begin{table}[tbp]
\footnotesize
\centering
\caption{Ranking of the feature importance according to $n$-fold cross validation of all 81 used metropolitan areas in the United States}
\label{feature-importance}
\input{feature_importance_tab.tex}
\end{table}
\end{scriptsize}

\end{center}

\subsection{Transforming the classification problem to a continuous prediction}

\begin{figure}[tbp]
\begin{center}
         \includegraphics[angle=0,scale=0.6]{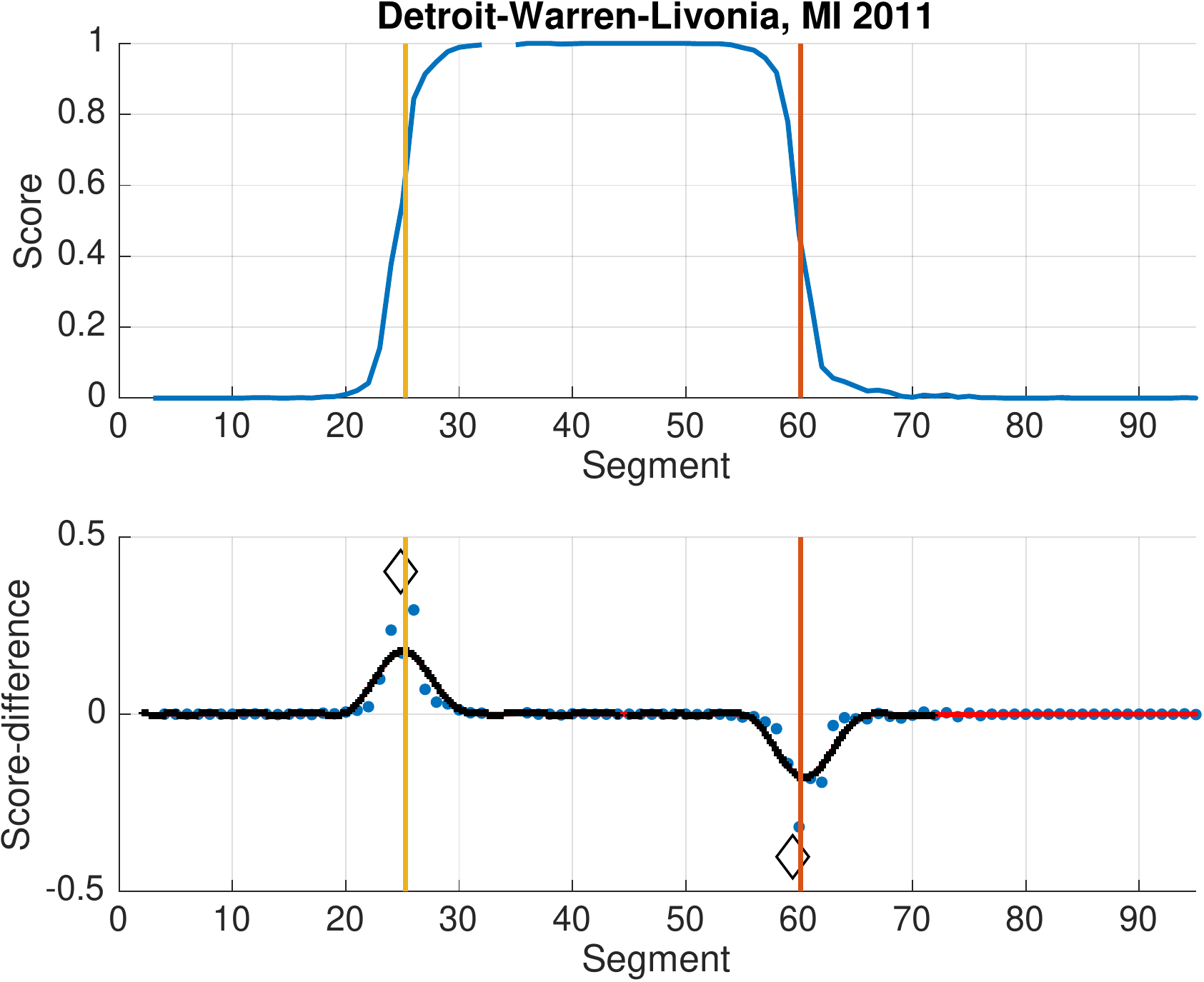}
\caption{Predicted score for Detroit-Warren-Livonia, Michigan in 2011 (top); conversion to a continuous time to wake and time to sleep prediction (bottom). Vertical lines indicate actual average ATUS traditional data, with diamonds showing the x-position of model predictions.}
 \label{detroit}
\end{center}
\end{figure}

The main focus of the prediction exercise was to up-scale a traditional time use survey using Internet data to derive sleep start and stop estimates in minutes world-wide. Using the $n$-fold cross-validated prediction, we predict each city and year combination based on all the other cities as input. Figure \ref{detroit} displays the independent segment predictions of sleep for the metropolitan area Detroit-Warren-Livonia, Michigan in 2011. The segments are shifted with the day starting at 4pm (Segment 64). The gap in the scoreline are the two segments after midnight, which are not used due the first and second difference of the fraction online input data.

The conversion to a continuous prediction follows the following procedure. First, the shifted prediction score is smoothed slightly with a parameter value of $0.06$, to remove the sharp edges of the independently drawn prediction as well as to interpolate the time gap after midnight. Shifting of the signal is of particular importance to accurately allow the smoothing algorithm to interpolate the gap. Second, the first difference of the smoothed signal is calculated as highlighted by the dots at the bottom of figure \ref{detroit}. Third, we use wavelet denoising (\textit{sym8}, level 1) to remove noise but at the same time preserve the sharp gradient from the $0$ to $1$ switch. Fourth, we use spline interpolation (black line) for the waking and to-sleep sections, separated at 3 am.  Finally, we calculate the x-position of the minimum and maximum of the two phases to derive the predicted start and stop times in minutes. The vertical lines represent the survey based start and stop estimates, while the diamonds mark our prediction outcome.

\begin{figure}[tbp]
\begin{center}
		
         \includegraphics[angle=0,scale=0.6]{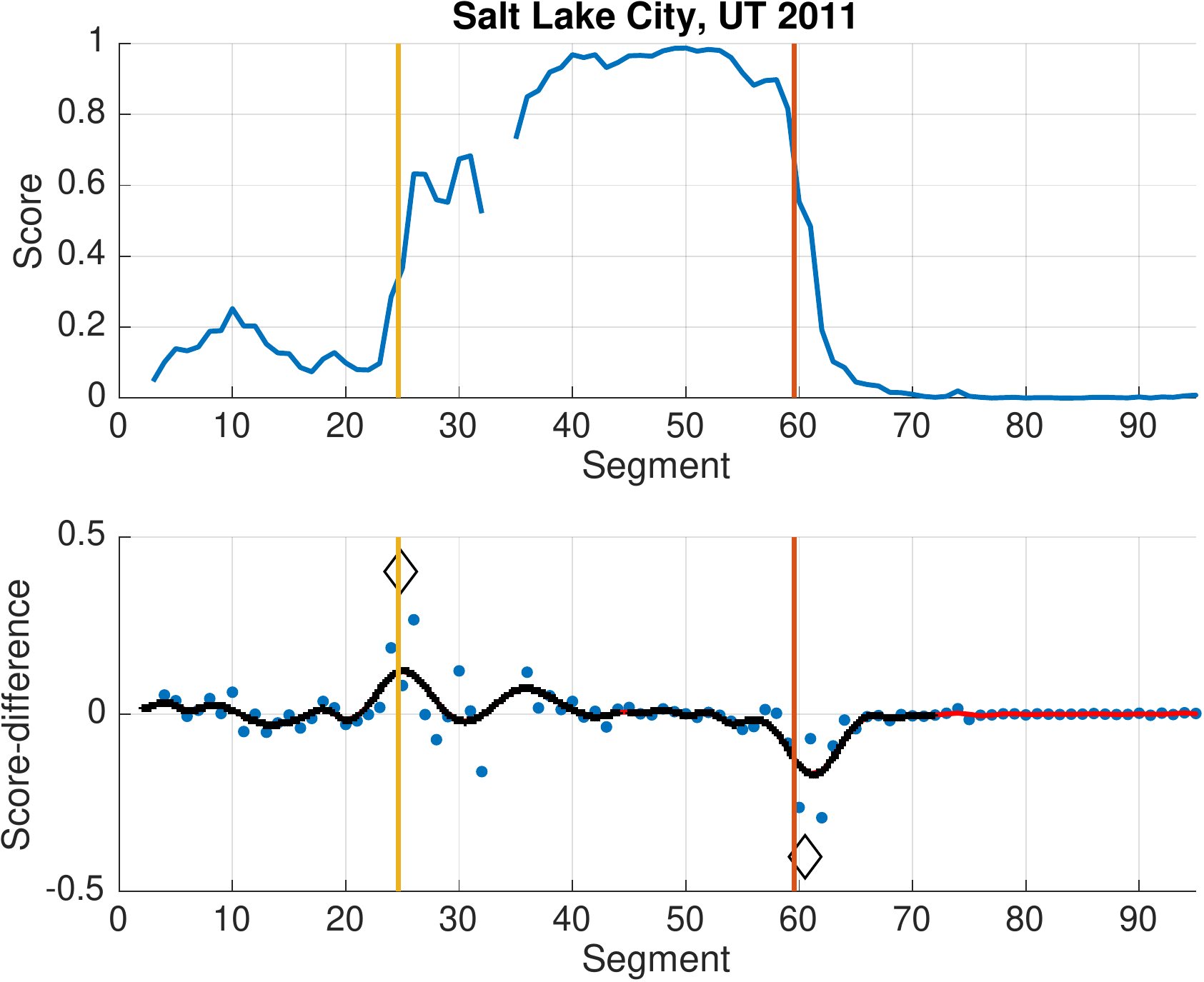}
         
\caption{Predicted score for Salt Lake City, Utah in 2011. }
\label{saltlake}
\end{center}
\end{figure}

\begin{figure}[tbp]
\begin{center}
         \includegraphics[angle=0,scale=0.6]{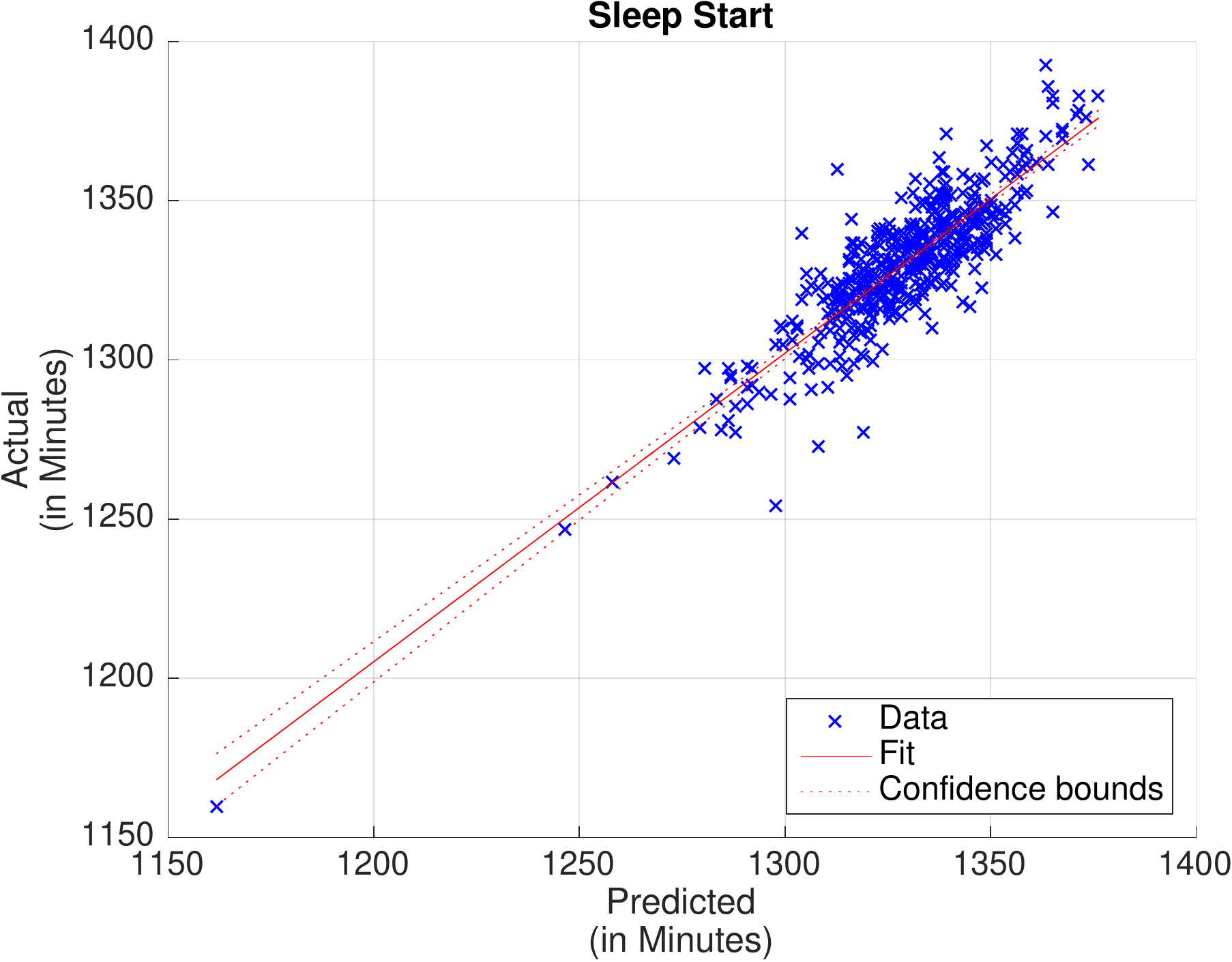}
\caption{Predicted sleep start time versus the actual sleep start time according to ATUS}
         \label{sleep_start}

\end{center}
\end{figure}

\begin{figure}[tbp]
\begin{center}
         \includegraphics[angle=0,scale=0.6]{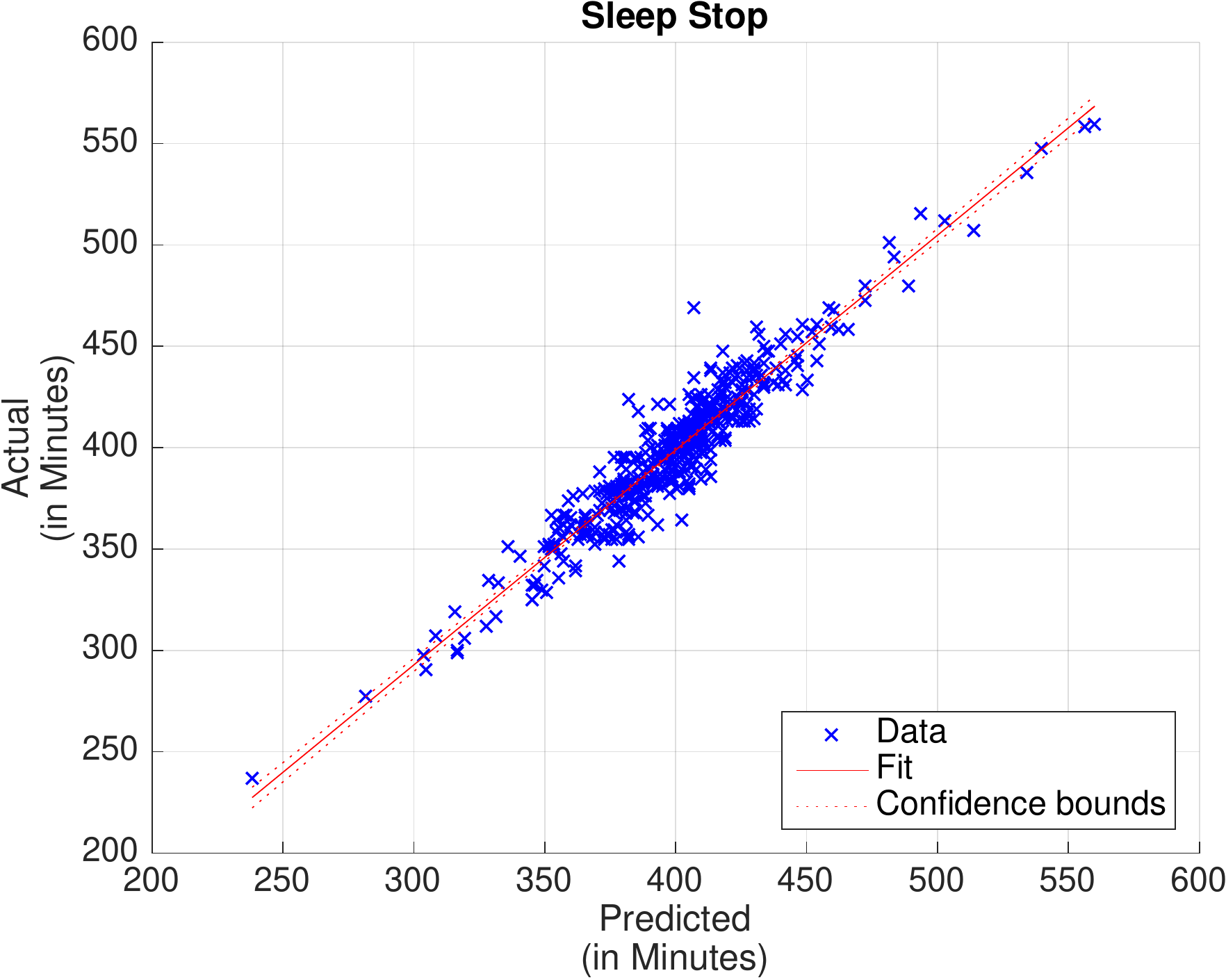}
\caption{Predicted sleep stop time versus the actual sleep start time according to ATUS}
         \label{sleep_stop}

\end{center}
\end{figure}

\begin{figure}[tbp]
\begin{center}
         \includegraphics[angle=0,scale=0.6]{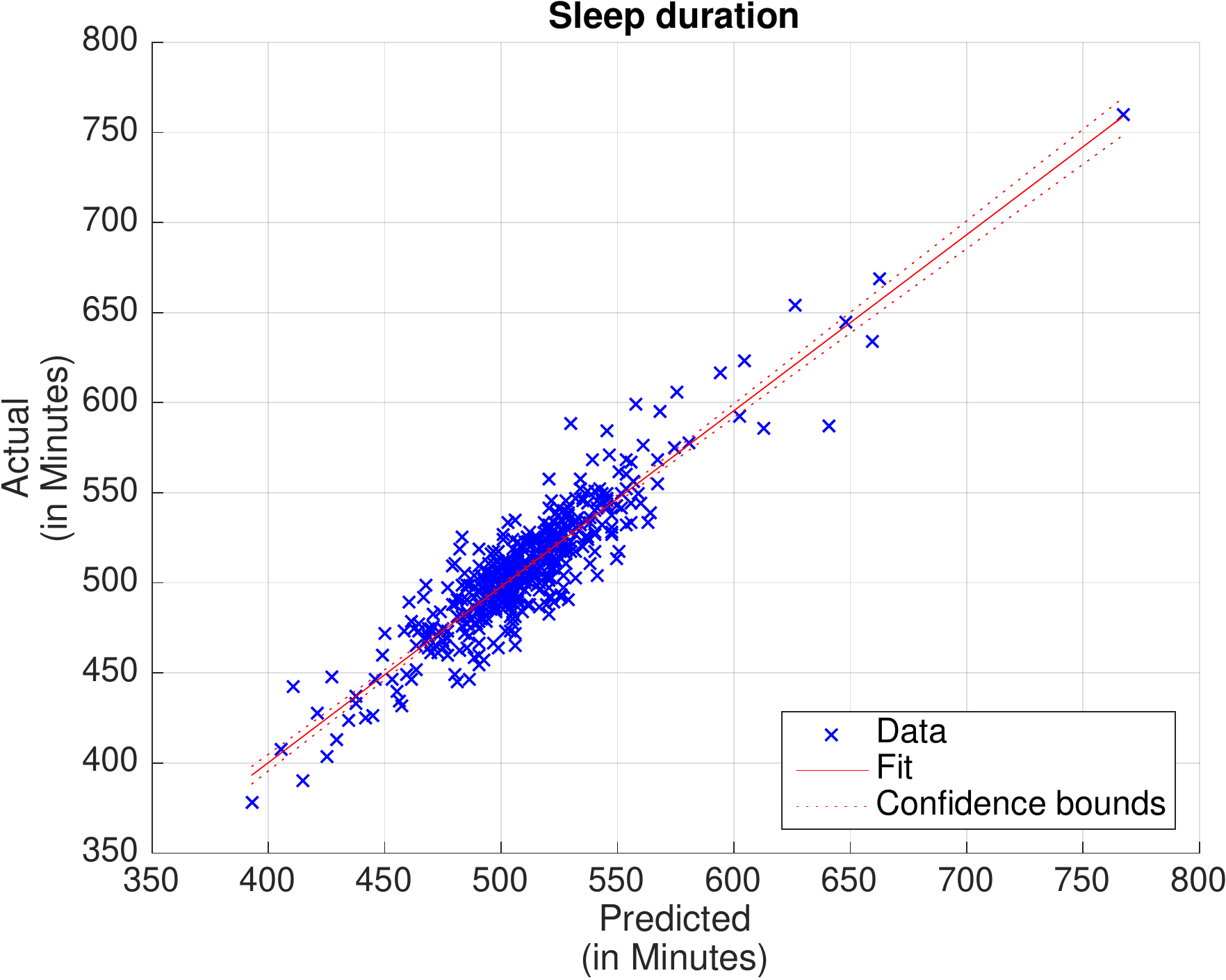}
\caption{Predicted sleep duration versus the actual sleep start time according to ATUS}
         \label{sleep_duration}

\end{center}
\end{figure}

The method was optimised to work well across all cities, including ones that have more varying sleep/wake prediction challenges such as can be seen in figure \ref{saltlake}. In general, the prediction surrounding the sleep stop time are more consistent in contrast to the sleep start time. We conjecture that this is an artefact of the standardised start time at workplaces and schools, while the to-sleep time has a higher variance. Figure \ref{sleep_start} shows the predicted sleep start time versus the actual start time according to the ATUS survey. The linear fit has a root mean squared error (RMSE) of $10.6$ minutes and a $R^2$ of $0.76$. On the other hand, the sleep stop time in figure \ref{sleep_stop}, we achieve an RMSE of $11.4$ minutes and a $R^2$ of $0.90$. The total estimate of sleep duration is calculated by the difference between stop and start time in figure \ref{sleep_duration}. The linear fit corresponds to an RMSE of $15.1$ and a $R^2$ of $0.83$.

\subsection{World wide prediction}

We train the model on all US metropolitan areas and use the previously described method for worldwide prediction. As we do not have confidence intervals or any other comparable prediction quality measures, we rely on the shape of the output signal. We enforce as a filter a rule that for a given city and year prediction at least 15 segments need to be above 0.9 and at least 20 segments below 0.1. We used visual inspection to tune these parameters to reduce the occurrence of predictions which did not follow the general patterns we observed in the cross-validation sample. Table \ref{sleep-tab} calculates the average by country across all cities and all years for each observations. We included only countries with at least 3 observations for the table. The least sleep duration is estimated for Japan followed by Slovenia, and the longest estimated for Spain and Argentina.

\begin{center}
\begin{scriptsize}
\input{sleep_tab.tex}

\end{scriptsize}
\end{center}

%
\section{Part 3: Internet penetration, incomes \& productivity}\label{sc:agg}

\subsection{Data}

The underlying data is an unbalanced panel dataset for 411 large regions from middle and high income countries for the years 2006-2012. The regions are defined by the OECD and normally correspond to the first subnational level (i.e. states in the United States or NUTS2 regions in Europe). Table \ref{tab.countries} presents a list of the countries and the respective number of regions per country in our data set.

\begin{center}
\newcolumntype{d}{D{.}{.}{3.3}}
\begin{small}
\begin{threeparttable}[htpb]
\caption{Countries and Number of Regions per Country}
\label{tab.countries}
\begin{tabular}{lccccc}
\noalign{\hrule height 1.5pt}
Country	&	\# of Regions	&	Country	&	\# of Regions	\\\hline 
Austria	&	9	&	Luxembourg	&	1	\\
Belgium	&	3	&	Mexico	&	16	\\
Brazil	&	22	&	Netherlands	&	4	\\
Canada	&	11	&	New Zealand	&	2	\\
Chile	&	4	&	Norway	&	7	\\
China	&	31	&	Poland	&	16	\\
Czech Republic	&	8	&	Portugal	&	6	\\
Denmark	&	5	&	Russia	&	71	\\
Estonia	&	1	&	Slovakia	&	4	\\
Finnland	&	1	&	Slovenia	&	2	\\
France	&	21	&	South Africa	&	4	\\
Germany	&	16	&	South Korea	&	6	\\
Greece	&	4	&	Spain	&	17	\\
Hungary	&	7	&	Sweden	&	8	\\
India	&	13	&	Switzerland	&	7	\\
Ireland	&	2	&	United Kingdom	&	12	\\
Italy	&	9	&	United States	&	51	\\
Japan	&	10	&						\\
\noalign{\hrule height 1.5pt}
\end{tabular}
\end{threeparttable}
\end{small}
\end{center}

Table \ref{tab.sum} presents the summary statistics for our key variables. The average regional GDP p.c. in our sample is \$ 27,330.02 which approximately corresponds to the regional GDP of Northeast England or Kyushu, Okinawa. Overall, the values range from \$1,591.07 (Uttar Pradesh, India) to \$ 174,423.4 (Washington, D.C., USA). 

The average IP per 1,000 inhabitants is 58.84 which is roughly the internet penetration in Conneticut or Lombardia. Average IP per 1,000 inhabitants ranges from 0.009 (Uttar Pradesh) to 1245.878 (Washington, D.C.). The IP data is aggregated and corrected for missoni bias as in Part 1, with the difference of spatial aggregation on regional\footnote{We are grateful to \textit{Monica Brezzi
 - Territorial Analysis and Statistics Unit,
OECD/GOV Regional Development Policy Division} for sharing the TL2 shapefile with us. } level as well as a minimum cut-off 200 IP addresses per month. We take the average over all months in a year to derive our IP measure.

To analyse the relationship between internet penetration and sectoral productivity we collected data on the Gross Value Added (per worker) for the following 7 sectors: Professional, technical, administrative Services and support Service activities; Public administrations, Education and Human Health; Information and Communication; Distributive Trade, Repairs, Transport, Hospitality; Industry (inc. Energy production); Other Services; Real Estate.

The classification of the sectors is based on the ISIC REV.4 (International Standard Industrial Classification of All Economic Activities, Rev.4) and represents the top level. The values in Table  \ref{tab.sum} are GVA per worker in USD at current prices and PPP.

\begin{center}
\newcolumntype{d}{D{.}{.}{3.3}}
\begin{small}
\begin{threeparttable}[htpb]
\caption{Descriptive Statistics}
\label{tab.sum}
\begin{tabular}{lccccc}
 
\noalign{\hrule height 1.5pt}
  Variable    &     Obs  &      Mean &   Std. Dev.   &    Min   &     Max\\\hline
 GDP p.c.    &   2,832  &  27330.02 &   18395.78  & 1591.07&   174423.40 \\
IP per Capita &   2,832  &  58.85 &    85.25 &   0.01  & 1245.88\\
Gross Value Added (GVA)\\
\quad Prof. Services   &    1,041 &      53030.00&     17231.70  &    11921.00 &    145291.00\\
\quad  Public. Admin.  &   1,220 &   43300.92   &  16957.10   &   11845.00  &    90795.00\\
\quad  Info \& COM  &   1,061 &   125771.3   &  51881.20   &   43426.00  &   352820.00\\
\quad  Trade \& Trans.  &   1,115 &   46958.33   & 12568.63   &   13149.00    &  95222.00\\
\quad  Industry  &   1,422 &   103357.8   & 69411.58   &   25410.00   &  732007.00\\
\quad  Other Services  &   1,061 &   87878.38   & 78601.96   &    4474.00   &  537219.00\\
\quad  Real Estate &   1,059 &   620622.9   & 783586.30   &  133105.00  &  5372430.00\\\noalign{\hrule height 1.5pt}
\end{tabular}
\end{threeparttable}
\end{small}
\end{center}

\subsection{Internet penetration and regional economic activity}

The objective of this analysis is to study the relationship between internet activity/penetration and GDP per capita at the region-level. For this purpose, we specify the following baseline regression model:

\begin{equation}\label{P.3_eq.1}
Y_{ict}=\alpha +\beta IP_{ict}+\lambda_t +\varepsilon_{ict}
\end{equation}
where $Y_{ict}$ is the natural log of GDP per capita in region $i$, country $c$ and year $t$, $IP_{ict}$ is the natural log of the number of active IP addresses per capita, $\alpha$ is a constant and $\beta$ is the coefficient of interest. $\lambda_t$ accounts for yearly shocks that are common to all regions (i.e. technological shocks). Using the natural log of GDP per capita as well as IP per capita allows us to directly interpret $\beta$ as an elasticity. Equation \ref{P.3_eq.1} is also the regression model underlying Figure 3 (panel a). 

The estimation results are presented in Table \ref{tab.gdp}, column 1. The estimated coefficient is positive and statistically significant at the 1\%-level. With respect to its magnitude, a 10\% increase in IP per capita is related to a 3.8 \% increase in regional GDP. p.c.

However, it is very likely that differences in both, the level of income, $Y$, and internet penetration, $IP$, are driven by time-invariant, region-specific characteristics. For example, regions in more developed countries have already a higher stock of telecommunications infrastructure that makes it easier to provide internet access. Region-specific history, culture and location can also jointly determine a region's level of development and adoption of the internet. Therefore, we augment equation \ref{P.3_eq.1} by a region-specific constant terms $\mu_{i}$ and estimate the following fixed effects (FE) model:

\begin{equation}
\label{P.3_eq.2}
Y_{ict}=\mu_{i}+\beta IP_{ict}+\lambda_t +\varepsilon_{ict}
\end{equation}

This approach is more reliable than including a set of other, observable covariates, because it enables us to account for all time-invariant factors that simultaneously influence regional development and regional internet penetration. The FE estimator not only exploits variation between regions but also variation within regions. In other words, we compare a region’s economic output in a year with low internet penetration to the economic output in a year with increased internet penetration. In practice, we estimate if region specific deviations in economic output are correlated with region-specific deviations in internet activity net of any positive and negative shocks that a country experiences in a year.

The estimation results are presented in Table \ref{tab.gdp}, column 2. Again, the coefficient is positive and statistically significant at the 1\% level. However, compared to column 1, the coefficient decreases by a factor 10 indicating that most of the variation comes from time-invariant, between regional differences.

Income and internet penetration could also follow some unobserved, regional trends which are, for example, the result of past or current changes in economic policy. To account for those underlying trends, we extend equation \ref{P.3_eq.2} by a vector of region specific time trends, $\tau_{i}t$. Further, there could be shocks that are common to all regions within a country (rather than all regions globally). For example, large scale infrastructure updates that improve access to the internet and also have positive impact on local GDP. To control for those effects, we further include vector of country-year specific dummy variables, $\theta_{ct}$.

\begin{equation}\label{P.3_eq.3}
Y_{ict}=\mu_{i}+\beta IP_{ict}+\theta_{ct}+\tau_{i}t  +\varepsilon_{ict}
\end{equation}

Basically, specification \ref{P.3_eq.3} analysis if region-specific deviations in GDP per capita are correlated to region-specific deviations in internet per capita net of any underlying, region-specific time trends and net of any country-year specific shocks.
 
Column (3) presents the results from estimating specification \ref{P.3_eq.3}. The coefficient suggests that a 10\% increase in IP per capita in a region is associated with a 0.8\% increase in GDP per capita. Again, one should be reminded that the estimated coefficient presents a correlation and does not allow for a causal interpretation. 

We then perform a number of robustness tests to check if the results are sensitive to changes in the lag structure or the composition of the sample. In column (4), we include internet penetration in year $t-1$ instead of the contemporaneous value and find that the coefficient is pretty similar. The specification in column (5) excludes observations with very high and very low (below 5\% and above 95\% of the distribution) values of regional GDP p.c. to check whether our results are driven by outliers with respect to regional income. Overall, the results are qualitatively the same. We then split the sample into observations with a regional GDP p.c. below (column 6) and above (7) the median GDP in the sample of U\$ 26,253. Again, the estimated coefficients are positive and statistically significant at the 1\% level. 

Interestingly, we find a somewhat stronger association between internet activity and regional development in regions below the median GDP as compared to regions above the median GDP. One could interpret those findings as an indication for potentially diminishing marginal returns of the internet.

Columns (8) and (9) present estimates excluding regions and years with very high and very low levels of internet penetration (below 5\% and above 95\% of the distribution) as well as observations from large countries (China, Russia and the USA), respectively. The baseline results do not appear to be sensitive to the exclusion of those observations. We also find that the correlation between GDP and IP is stronger in relatively poorer regions and for the years prior to 2008 (Table E, Supp. Material). 

Finally, we split the sample into observations prior to 2008 (column 10) and post 2008 (column 11). The motivation behind this exercise is to account for the introduction of smartphones in 2008 which led to a rise in mobile internet and therefore an overall rise in internet per capita. We find that internet penetration and GDP per capita are again positively correlated. However, the correlation seems to be stronger in the period up to 2008. This could be due to the increased use of the internet for leisure activities and private communication as a result of the diffusion of smartphones. 

Overall, this exercise shows that there is a strong and robust relationship between internet penetration and GDP per capita. This provides evidence that high resolution internet data can also be used, at a more aggregate level (in this case at a regional and yearly level), to measure more complex forms of human activity such as overall economic development. Finally, it is important to point out that the estimated coefficients are correlations and do not allow causal inference between internet activity and regional economic development. Providing causal estimates of the impact of internet activity on local economic development is subject to an ongoing research project by the authors.

 \newpage 
\renewcommand{\arraystretch}{1}
\begin{landscape}
\begin{center}
\newcolumntype{d}{D{.}{.}{3.3}}
\begin{small}
\begin{threeparttable}[htpb]
\caption{Internet Penetration and Regional GDP}
\label{tab.gdp}
\begin{tabular}{p{1.7cm}ddddddddddd}

\noalign{\hrule height 1.5pt}
 & \multicolumn{1}{c}{1}& \multicolumn{1}{c}{2} & \multicolumn{1}{c}{3}& \multicolumn{1}{c}{4}& \multicolumn{1}{c}{5}& \multicolumn{1}{c}{6}& \multicolumn{1}{c}{7}& \multicolumn{1}{c}{8}& \multicolumn{1}{c}{9}& \multicolumn{1}{c}{10}& \multicolumn{1}{c}{11}\\
  & \multicolumn{1}{c}{OLS} & \multicolumn{1}{c}{Region} & \multicolumn{1}{c}{Base}&\multicolumn{1}{c}{1$^{st}$ Lag}&\multicolumn{1}{c}{Exc. Hi \& Lo}&\multicolumn{1}{c}{Below}&\multicolumn{1}{c}{Above}&\multicolumn{1}{c}{Exc. Hi \& Lo}&\multicolumn{1}{c}{Exc. CHN,}&\multicolumn{1}{c}{Year}&\multicolumn{1}{c}{Year}\\
    & \multicolumn{1}{c}{} & \multicolumn{1}{c}{  FEs} & \multicolumn{1}{c}{ }&\multicolumn{1}{c}{} &\multicolumn{1}{c}{GDP }&\multicolumn{1}{c}{Median GDP}&\multicolumn{1}{c}{Median GDP}&\multicolumn{1}{c}{IP p.c.}&\multicolumn{1}{c}{RUS, USA}&\multicolumn{1}{c}{$\leq$ 2008}&\multicolumn{1}{c}{$>$ 2008}\\\hline
 $Ln(IP)_{ict}$ & 0.377*** & 0.031*** & 0.080*** &  & 0.054*** & 0.074*** & 0.040*** & 0.075*** & 0.057*** & 0.072*** & 0.042*** \\
 & (0.006) & (0.006) & (0.011) &  & (0.010) & (0.017) & (0.010) & (0.014) & (0.009) & (0.014) & (0.008) \\
$Ln(IP)_{ict-1}$ &  &  &  & 0.072*** \\
 &  &  &  & (0.011) \\\hline
$\mu_i$ & \multicolumn{1}{c}{No}& \multicolumn{1}{c}{Yes}& \multicolumn{1}{c}{Yes}& \multicolumn{1}{c}{Yes}& \multicolumn{1}{c}{Yes}& \multicolumn{1}{c}{Yes}& \multicolumn{1}{c}{Yes}& \multicolumn{1}{c}{Yes}& \multicolumn{1}{c}{Yes}& \multicolumn{1}{c}{Yes}& \multicolumn{1}{c}{Yes}	\\
$\lambda_t$	& \multicolumn{1}{c}{Yes}& \multicolumn{1}{c}{No}& \multicolumn{1}{c}{No}& \multicolumn{1}{c}{No}& \multicolumn{1}{c}{No}& \multicolumn{1}{c}{No}& \multicolumn{1}{c}{No}& \multicolumn{1}{c}{No}& \multicolumn{1}{c}{No}& \multicolumn{1}{c}{No}& \multicolumn{1}{c}{No}	\\
$\theta_{ct}$	 & \multicolumn{1}{c}{No}& \multicolumn{1}{c}{No}& \multicolumn{1}{c}{Yes}& \multicolumn{1}{c}{Yes}	& \multicolumn{1}{c}{Yes}& \multicolumn{1}{c}{Yes}& \multicolumn{1}{c}{Yes}& \multicolumn{1}{c}{Yes}& \multicolumn{1}{c}{Yes}& \multicolumn{1}{c}{Yes}& \multicolumn{1}{c}{Yes}\\
$\tau_i t$	 & \multicolumn{1}{c}{No}& \multicolumn{1}{c}{No}& \multicolumn{1}{c}{Yes}& \multicolumn{1}{c}{Yes}& \multicolumn{1}{c}{Yes}& \multicolumn{1}{c}{Yes}& \multicolumn{1}{c}{Yes}& \multicolumn{1}{c}{Yes}& \multicolumn{1}{c}{Yes}& \multicolumn{1}{c}{Yes}& \multicolumn{1}{c}{Yes}	\\
Observations & \multicolumn{1}{c}{2,832} & \multicolumn{1}{c}{2,832} & \multicolumn{1}{c}{2,832}& \multicolumn{1}{c}{2,421} & \multicolumn{1}{c}{2,550}  & \multicolumn{1}{c}{1,416} & \multicolumn{1}{c}{1,416} & \multicolumn{1}{c}{2,549} & \multicolumn{1}{c}{1,761} & \multicolumn{1}{c}{1,205} & \multicolumn{1}{c}{1,627} \\
 Number of regions & \multicolumn{1}{c}{ 411} &\multicolumn{1}{c}{ 411 }& \multicolumn{1}{c}{411} & \multicolumn{1}{c}{389 }&\multicolumn{1}{c}{ 221 }& \multicolumn{1}{c}{218} & \multicolumn{1}{c}{403 }& \multicolumn{1}{c}{258 }& \multicolumn{1}{c}{411} & \multicolumn{1}{c}{411 }\\ \hline

$R^2$ &\multicolumn{1}{c}{ 0.605} & \multicolumn{1}{c}{0.546} &\multicolumn{1}{c}{0.986}  &\multicolumn{1}{c}{0.990} &\multicolumn{1}{c}{0.989}&\multicolumn{1}{c}{0.972 }&\multicolumn{1}{c}{ 0.972}&\multicolumn{1}{c}{ 0.985}&\multicolumn{1}{c}{0.991 }&\multicolumn{1}{c}{0.996 }&\multicolumn{1}{c}{0.996 }

\\\noalign{\hrule height 1.5pt}
\end{tabular}
\begin{tablenotes}
\item \textit{Notes:} Dependent variable is the natural log of GDP per capita in region $i$ and year $t$. Robust Standard Errors in parentheses. $^{***}$ , $^{**}$, $^{*}$ indicate significance at the 1, 5 and 10\%-level, respectively. 
\end{tablenotes}
\end{threeparttable}
\end{small}
\end{center}
\end{landscape}
\newpage

\subsection{Sectoral effects of internet activity}
Figure 4 B  presents the correlation coefficients between internet per capita and Gross Value added per worker at the regional level for seven different industries.

In this section, we explain the empirical analysis behind those results. In general, Gross Value Added (GVA) per worker can be used to compare productivity between different sectors. Again to perform this analysis we had to rely on limited publicly available data.  The raw data stems again from the OECD, but compared to data on regional GDP, data on sectoral GVA is only available for a shorter time period. The number of observations ranges between 1,041 and 1,422 (as compared to 2,832 in the regional income analysis).

Whilst the International Standard Industrial Classification of All Economic Activities (ISIC), Rev. 4\footnote{See \url{http://unstats.un.org/unsd/cr/registry/isic-4.asp}.} provides a total of 21 `sections', the OECD raw data on GVA merges several of these sections together, leaving a total of 11 single, or merged, `sectors' for analysis.

The seven resultant `sectors' presented in Figure 4(B) of the main paper, representing a total of 13 underlying sections in ISIC Rev.4 are as follows:
\begin{enumerate}
\item Prof \& Admin Support services:  ISIC Rev.4 section {\bf N}:Administrative and support service activities. This sector includes rental and leasing of non-real estate goods; employment agencies; travel, tour and reservation services; security and investigation services; cleaning and facility support services; call centres, photocopying and packaging services.
\item Public Admin, Education, Health: OECD raw data, `public administration, compulsory social services, education, human health', combining ISIC Rev.4 sections {\bf O}: Public administration and defence, compulsory social security; {\bf P}:Education; and {\bf Q}:Human health and social work activities. This sector includes activities such as administration of the state, foreign affairs, defence, public order/safety, social security; pre-primary, primary, secondary, vocational, tertiary, and cultural education; hospital activities, medical and dental activities, residential care, mental health, social work activities.
\item Information \& Communication: ISIC Rev.4 section {\bf J}:Information \& Communication. This sector includes publishing, motion picture, video and television production, sound recording, music publishing, radio and television programming and broadcast, telecommunications, computer programming consultancy and related activities, data processing, hosting, web portals and news agency activities.
\item Trade repairs, Transport, Hosp.: OECD Raw data,  `distributive trade, repairs, transport, accommodation, food service activities', combining ISIC Rev.4 sections {\bf G}:Wholesale and retail trade, repair of motor vehicles and motorcycles; {\bf H}:Transportation and storage; and {\bf I}:Accommodation and food service activities. This sector includes the sale, maintenance and repair of motor vehicles and motorcycles; wholesale and retail across all product domains; land, water and air transportation; warehousing, cargo handling; accommodation, restaurants and event catering.
\item Industry (inc. Energy): OECD raw data, `Industry including energy', combining ISIC Rev.4 sections {\bf B}:Mining and quarrying; {\bf D}:Electricty, gas, steam and air conditioning supply; and {\bf E}:Water supply, sewerage, waste management and remediation activities. This sector includes mining and extraction in all forms; power generation, transmission and distribution; water collection, treatment and supply, sewerage, waste collection.
\item Other personal services: ISIC Rev.4 section {\bf S}:Other service activities. This sector includes membership organisations (unions, religious, professional); computer, household good, and personal good repair; washing, hairdressing, funerals.
\item Real Estate: ISIC Rev.4 section {\bf L}:Real estate activities.
\end{enumerate}

We follow a similar empirical strategy outlined in equation \ref{P.3_eq.2}, where we account for unobserved, time-invariant factors by including a vector of region specific fixed effects, $\mu_{i}$. Due to the short period of data coverage, we are unable to include region specific time trends. However, we account for unobserved contemporaneous shocks (i.e. technological innovations) that are common to all regions by including a vector of year dummies, $\lambda_t$.

For each sector, we estimate the following equation:
\begin{equation}\label{P.3_eq.4}
GVA_{ict}=\mu_{i}+\gamma IP_{ict}+\lambda_t +u_{ict}
\end{equation}

The dependent variable $ GVA_{ict}$ is the natural log of the sectoral Gross Value Added per worker in current U\$ (PPP-adjusted). Although we performed estimates for all 11 sectors we have GVA data on, we only find a statistically significant correlation between internet penetration and productivity for seven sectors. We do not find a systematic relationship between those two variables for Agriculture, Mining, Construction, and Finance. 

The results are presented in Table \ref{tab.sectoral}. Overall, we find that the relationship between increased internet activity and sectoral GVA varies widely between sectors. On the one hand, columns 1 to 3 reveal that an increase in regional internet activity is associated with a decrease in regional GVA per worker in Professional Support Services, Public Administration as well as Information and Communications. On the other hand, the results in columns 4 to 7 show that higher internet penetration is positively associated with GVA per worker in Distributive Trade and Transport, Industry, Other Services and Real Estate.

These results reveal an interesting relationship between increased internet penetration and sectoral change. In some sectors the internet has changed consumption patterns and individual households decreased consumption of those services from local suppliers in favour of consuming services directly form the internet. Publishing activities, motion pictures, computer programming (all part of Information \& Communication sector) or accounting and advertising services (part of Professional Support Services) are an example of those shifts. In contrast, the internet related changes in consumption led to an increase in demand for other local services such as transport. In addition, increased internet penetration might have enabled local producers and service providers to outsource certain activities and decrease labour costs. Once again, the estimates are based on correlations and should not be interpreted as causal effects.

\newpage

\renewcommand{\arraystretch}{1}
\begin{landscape}
\begin{center}
\newcolumntype{d}{D{.}{.}{3.3}}
\begin{small}
\begin{threeparttable}[htpb]
\caption{Sectoral Effects - Internet Penetration and Gross-Value-Added (GVA) in Each Sector}
\label{tab.sectoral}
\begin{tabular}{lddddddd}

\noalign{\hrule height 1.5pt}
 & \multicolumn{1}{c}{1}& \multicolumn{1}{c}{2} & \multicolumn{1}{c}{3}& \multicolumn{1}{c}{4}& \multicolumn{1}{c}{5}& \multicolumn{1}{c}{6}& \multicolumn{1}{c}{7}\\
  & \multicolumn{1}{c}{\textbf{Prof. \& Admin. }}& \multicolumn{1}{c}{\textbf{Public Admin., }}& \multicolumn{1}{c}{\textbf{Information \& }}& \multicolumn{1}{c}{\textbf{Trade, Repairs}}& \multicolumn{1}{c}{\textbf{Industry}}& \multicolumn{1}{c}{\textbf{Other}}& \multicolumn{1}{c}{\textbf{Real }}\\
  
  & \multicolumn{1}{c}{\textbf{Support Services }}& \multicolumn{1}{c}{\textbf{Edu., Health }}& \multicolumn{1}{c}{\textbf{Communications }}& \multicolumn{1}{c}{\textbf{Transport, Hosp. }}& \multicolumn{1}{c}{\textbf{  }}& \multicolumn{1}{c}{\textbf{Services }}& \multicolumn{1}{c}{\textbf{Estate}}\\\hline

$Ln(IPperCapita)_{ict}$ & -0.031*** & -0.015*** & -0.015** & 0.009** & 0.010* & 0.024*** & 0.051*** \\
 & (0.006) & (0.004) & (0.007) & (0.004) & (0.006) & (0.006) & (0.010) \\

\hline

Observations & \multicolumn{1}{c}{1,041} & \multicolumn{1}{c}{1,220} & \multicolumn{1}{c}{1,061 } & \multicolumn{1}{c}{1,115}& \multicolumn{1}{c}{1,422}& \multicolumn{1}{c}{1,061 }& \multicolumn{1}{c}{1,059 }\\
Number of regions & \multicolumn{1}{c}{ 169} & \multicolumn{1}{c}{204} & \multicolumn{1}{c}{ 179 } & \multicolumn{1}{c}{189 }& \multicolumn{1}{c}{242}& \multicolumn{1}{c}{ 179 }& \multicolumn{1}{c}{ 179 }\\
$R^2$ &\multicolumn{1}{c}{0.385} & \multicolumn{1}{c}{0.704 } &\multicolumn{1}{c}{0.284}  &\multicolumn{1}{c}{ 0.598} &\multicolumn{1}{c}{ 0.550 } &\multicolumn{1}{c}{0.387 } &\multicolumn{1}{c}{ 0.415}  \\\noalign{\hrule height 1.5pt}
\end{tabular}
\begin{tablenotes}
\item \textit{Notes:} Dependent variable is the Gross Value Added (GVA) in the respective sector. Values are the logged GVA in USD per worker at current prices and PPP.  All specifications include a region-specific constant term, $\mu_i$, and a year specific term, $\lambda_t$.  Robust Standard Errors in parentheses. $^{***}$ , $^{**}$, $^{*}$ indicate significance at the 1, 5 and 10\%-level, respectively. 
\end{tablenotes}
\end{threeparttable}
\end{small}
\end{center}
\end{landscape}
\newpage

\subsection{List of regions}

(AUT)	Burgenland,
	Lower Austria,
	Vienna,
	Carinthia,
	Styria,
	Upper Austria,
	Salzburg,
	Tyrol,
	Vorarlberg,
(BEL)	Brussels Capital Region,
	Flemish Region,
	Wallonia,
(BRA)	Amazonas,
	Pará,
	Rondônia,
	Alagoas,
	Bahia,
	Ceará,
	Maranhão,
	Paraíba,
	Pernambuco,
	Piauí,
	Rio Grande Do Norte,
	Espírito Santo,
	Minas Gerais,
	Rio De Janeiro,
	São Paulo,
	Paraná,
	Rio Grande Do Sul,
	Santa Catarina,
	Distrito Federal,
	Goiás,
	Mato Grosso,
	Mato Grosso Do Sul,
(CAN)	Newfoundland and Labrador,
	Prince Edward Island,
	Nova Scotia,
	New Brunswick,
	Quebec,
	Ontario,
	Manitoba,
	Saskatchewan,
	Alberta,
	British Columbia,
	Northwest Territories,
(CHE)	Lake Geneva Region,
	Espace Mittelland,
	Northwestern Switzerland,
	Zurich,
	Eastern Switzerland,
	Central Switzerland,
	Ticino,
(CHL)	Valparaíso,
	Bío-Bío,
	Santiago Metropolitan,
	Los Rios,
(CHN)	Beijing,
	Tianjin,
	Hebei,
	Shanxi,
	Inner Mongolia,
	Liaoning,
	Jilin,
	Heilongjiang,
	Shanghai,
	Jiangsu,
	Zhejiang,
	Anhui,
	Fujian,
	Jiangxi,
	Shandong,
	Henan,
	Hubei,
	Hunan,
	Guangdong,
	Guangxi,
	Hainan,
	Chongqing,
	Sichuan,
	Guizhou,
	Yunnan,
	Tibet,
	Shaanxi,
	Gansu,
	Qinghai,
	Ningxia,
	Xinjiang,
(CZE)	Prague,
	Central Bohemian Region,
	Southwest,
	Northwest,
	Northeast,
	Southeast,
	Central Moravia,
	Moravia-Silesia,
(DEU)	Baden-Württemberg,
	Bavaria,
	Berlin,
	Brandenburg,
	Bremen,
	Hamburg,
	Hesse,
	Mecklenburg-Vorpommern,
	Lower Saxony,
	North Rhine-Westphalia,
	Rhineland-Palatinate,
	Saarland,
	Saxony,
	Saxony-Anhalt,
	Schleswig-Holstein,
	Thuringia,
(DNK)	Capital,
	Zealand,
	Southern Denmark,
	Central Jutland,
	Northern Jutland,
(EST)	Estonia,
(ESP)	Galicia,
	Asturias,
	Cantabria,
	Basque Country,
	Navarra,
	La Rioja,
	Aragon,
	Madrid,
	Castile and León,
	Castile-La Mancha,
	Extremadura,
	Catalonia,
	Valencia,
	Balearic Islands,
	Andalusia,
	Murcia,
	Canary Islands,
(FIN)	Western Finland,
(FRA)	Ile-de-France,
	Champagne-Ardenne,
	Picardy,
	Upper Normandy,
	Centre-Val de Loire,
	Lower Normandy,
	Burgundy,
	Nord-Pas-de-Calais,
	Lorraine,
	Alsace,
	Franche-Comté,
	Pays de la Loire,
	Brittany,
	Poitou-Charentes,
	Aquitaine,
	Midi-Pyrénées,
	Limousin,
	Rhône-Alpes,
	Auvergne,
	Languedoc-Roussillon,
	Provence-Alpes-Côte d'Azur,
(GRC)	Northern Greece,
	Central Greece,
	Athens,
	Aegean Islands and Crete,
(HUN)	Central Hungary,
	Central Transdanubia,
	Western Transdanubia,
	Southern Transdanubia,
	Northern Hungary,
	Northern Great Plain,
	Southern Great Plain,
(IRL)	Border, Midland and Western,
	Southern and Eastern,
(IND)	National Capital Territory of Delhi,
	Uttar Pradesh,
	West Bengal,
	Gujarat,
	Maharashtra,
	Kerala,
	Chandigarh,
	Haryana,
	Orissa,
	Madhya Pradesh,
	Andhra Pradesh,
	Karnataka,
	Tamil Nadu,
(ITA)	Piedmont,
	Liguria,
	Lombardy,
	Abruzzo,
	Campania,
	Apulia,
	Calabria,
	Sicily,
	Sardinia,
(JPN)	Hokkaido,
	Tohoku,
	Northern-Kanto, Koshin,
	Southern-Kanto,
	Hokuriku,
	Toukai,
	Kansai region,
	Chugoku,
	Shikoku,
	Kyushu, Okinawa,
(KOR)	Capital Region,
	Gyeongnam Region,
	Gyeongbuk Region,
	Jeolla Region,
	Chungcheong Region,
	Gangwon Region,
(LUX)	Luxembourg,
(MEX)	Baja California Norte,
	Colima,
	Chihuahua,
	Federal District,
	Guanajuato,
	Jalisco,
	Mexico,
	Michoacan,
	Morelos,
	Nuevo Leon,
	Puebla,
	Queretaro,
	Quintana Roo,
	Sinaloa,
	Sonora,
	Veracruz,
(NLD)	North Netherlands,
	East Netherlands,
	West Netherlands,
	South Netherlands,
(NOR)	Oslo and Akershus,
	Hedmark and Oppland,
	South-Eastern Norway,
	Agder and Rogaland,
	Western Norway,
	Trøndelag,
	Northern Norway,
(NZL)	North Island,
	South Island,
(POL)	Lodzkie,
	Mazovia,
	Lesser Poland,
	Silesia,
	Lublin Province,
	Podkarpacia,
	Swietokrzyskie,
	Podlasie,
	Greater Poland,
	West Pomerania,
	Lubusz,
	Lower Silesia,
	Opole region,
	Kuyavian-Pomerania,
	Warmian-Masuria,
	Pomerania,
(PRT)	North,
	Algarve,
	Central Portugal,
	Lisbon,
	Alentejo,
	Madeira,
(RUS)	Belgorod Oblast,
	Bryansk Oblast,
	Vladimir Oblast,
	Voronezh Oblast,
	Ivanovo Oblast,
	Kaluga Oblast,
	Kostroma Oblast,
	Kursk Oblast,
	Lipetsk Oblast,
	Moscow Oblast,
	Oryol Oblast,
	Ryazan Oblast,
	Smolensk Oblast,
	Tambov Oblast,
	Tver Oblast,
	Tula Oblast,
	Yaroslavl Oblast,
	City of Moscow,
	Republic of Karelia,
	Komi republic,
	Arkhangelsk Oblast,
	Vologda Oblast,
	Kaliningrad Oblast,
	Murmansk Oblast,
	Novgorod Oblast,
	Pskov Oblast,
	Federal City of Saint Petersburg,
	Krasnodar Krai,
	Astrakhan Oblast,
	Volgograd Oblast,
	Rostov Oblast,
	Republic of Dagestan,
	Republic of North Ossetia-Alania,
	Stavropol Krai,
	Republic of Bashkorstostan,
	Mari El Republic,
	Republic of Mordovia,
	Republic of Tatarstan,
	Udmurt Republic,
	Chuvash Republic,
	Perm Krai,
	Kirov Oblast,
	Nizhny Novgorod Oblast,
	Orenburg Oblast,
	Penza Oblast,
	Samara Oblast,
	Saratov Oblast,
	Ulianov Oblast,
	Kurgan Oblast,
	Sverdlovsk Oblast,
	Tyumen Oblast,
	Khanty-Mansi Autonomous Okrug - Yugra,
	Yamalo-Nenets Autonomous Okrug,
	Chelyabinsk Oblast,
	Buryat Republic,
	Republic of Khakassia,
	Altai Krai,
	Zabaykalsky Krai,
	Krasnoyarsk Krai,
	Irkutsk Oblast,
	Kemerovo oblast,
	Novosibirsk Oblast,
	Omsk Oblast,
	Tomsk Oblast,
	Sakha Republic (Yakutia),
	Kamchatka Krai,
	Primorsky Krai,
	Khabarovsk Krai,
	Amur Oblast,
	Magadan Oblast,
	Sakhalin Oblast,
(SWE)	Stockholm,
	East Middle Sweden,
	Småland with Islands,
	South Sweden,
	West Sweden,
	North Middle Sweden,
	Central Norrland,
	Upper Norrland,
(SVN)	Eastern Slovenia,
	Western Slovenia,
(SVK)	Bratislava Region,
	West Slovakia,
	Central Slovakia,
	East Slovakia,
(GBR)	North East England,
	North West England,
	Yorkshire and The Humber,
	East Midlands,
	West Midlands,
	East of England,
	Greater London,
	South East England,
	South West England,
	Wales,
	Scotland,
	Northern Ireland,
(USA)	Alabama,
	Alaska,
	Arizona,
	Arkansas,
	California,
	Colorado,
	Connecticut,
	Delaware,
	District of Columbia,
	Florida,
	Georgia,
	Hawaii,
	Idaho,
	Illinois,
	Indiana,
	Iowa,
	Kansas,
	Kentucky,
	Louisiana,
	Maine,
	Maryland,
	Massachusetts,
	Michigan,
	Minnesota,
	Mississippi,
	Missouri,
	Montana,
	Nebraska,
	Nevada,
	New Hampshire,
	New Jersey,
	New Mexico,
	New York,
	North Carolina,
	North Dakota,
	Ohio,
	Oklahoma,
	Oregon,
	Pennsylvania,
	Rhode Island,
	South Carolina,
	South Dakota,
	Tennessee,
	Texas,
	Utah,
	Vermont,
	Virginia,
	Washington,
	West Virginia,
	Wisconsin,
	Wyoming,
(ZAF)	Eastern Cape,
	Gauteng,
	KwaZulu-Natal,
	Western Cape,

\newpage
\bibliography{appendix}
\bibliographystyle{plain}
\end{document}

%% file: ip_cap_table.tex
\begin{longtable}{lrrrrrrrrr}
\caption{IP per capita rankings and diffusion parameters}\label{ip-cap-tab} \\
\noalign{\hrule height 1.5pt}
Country & \multicolumn{1}{l}{\begin{tabular}[c]{@{}l@{}}Number \\ of Cities\end{tabular}} & \multicolumn{1}{l}{\begin{tabular}[c]{@{}l@{}}Rank \\ 2006\end{tabular}} & \multicolumn{1}{l}{\begin{tabular}[c]{@{}l@{}}Rank \\ 2012\end{tabular}} & \multicolumn{1}{l}{Change} & \multicolumn{1}{l}{\begin{tabular}[c]{@{}l@{}}IP per capita \\ level 2012\end{tabular}} & \multicolumn{1}{l}{\begin{tabular}[c]{@{}l@{}}Asymptotic\\  Limit\end{tabular}} & \multicolumn{1}{l}{\begin{tabular}[c]{@{}l@{}}Growth\\  Rate\end{tabular}} & \multicolumn{1}{l}{Year 1\%} & \multicolumn{1}{l}{Year 99\%} \\ \hline
Germany & 32 & 7 & 1 & 6 & Saturated & 0.481 & 0.083 & 2004 & 2013 \\
Denmark & 4 & 2 & 2 & 0 & Saturated & 0.318 & 0.113 & 2003 & 2010 \\
Estonia & 1 & 4 & 3 & 1 & Saturated & 0.350 & 0.084 & 2003 & 2012 \\
South Korea & 9 & 22 & 4 & 18 & Saturated & 0.575 & 0.044 & 2003 & 2021 \\
Norway & 3 & 1 & 5 & -4 & High & 0.320 & 0.216 & 2004 & 2008 \\
Switzerland & 7 & 14 & 6 & 8 & High & 0.921 & 0.032 & 2003 & 2026 \\
Spain & 16 & 28 & 7 & 21 & High & 0.632 & 0.038 & 2003 & 2023 \\
Macedonia & 1 & 51 & 8 & 43 & High & 0.325 & 0.074 & 2005 & 2016 \\
Slovakia & 2 & 23 & 9 & 14 & High & 0.284 & 0.057 & 2003 & 2016 \\
Slovenia & 2 & 5 & 10 & -5 & High & 0.242 & 0.100 & 2003 & 2011 \\
Sweden & 10 & 10 & 11 & -1 & High & 0.311 & 0.047 & 2001 & 2017 \\
Latvia & 1 & 12 & 12 & 0 & High & 0.563 & 0.023 & 1997 & 2030 \\
Hungary & 2 & 18 & 13 & 5 & High & 0.245 & 0.129 & 2005 & 2011 \\
Iceland & 1 & 3 & 14 & -11 & High & 0.196 & 0.133 & 2004 & 2009 \\
Netherlands & 10 & 9 & 15 & -6 & High & 0.220 & 0.057 & 2001 & 2014 \\
Ireland & 2 & 20 & 16 & 4 & High & 0.204 & 0.088 & 2003 & 2012 \\
France & 28 & 8 & 17 & -9 & High & 0.255 & 0.039 & 1998 & 2018 \\
Saudi Arabia & 1 & 76 & 18 & 58 & High & 0.490 & 0.094 & 2009 & 2017 \\
United States & 70 & 6 & 19 & -13 & High & 0.182 & 0.104 & 2003 & 2010 \\
Bulgaria & 5 & 41 & 20 & 21 & High & 0.316 & 0.042 & 2003 & 2021 \\
Finland & 4 & 16 & 21 & -5 & High & 0.289 & 0.035 & 2000 & 2022 \\
Croatia & 1 & 43 & 22 & 21 & High & 0.646 & 0.035 & 2004 & 2026 \\
Taiwan & 3 & 25 & 23 & 2 & High & 0.266 & 0.036 & 2000 & 2022 \\
Italy & 21 & 36 & 24 & 12 & High & 0.227 & 0.042 & 2001 & 2020 \\
Portugal & 3 & 42 & 25 & 17 & High & 0.161 & 0.180 & 2006 & 2011 \\
Cyprus & 2 & 37 & 26 & 11 & High & 0.236 & 0.166 & 2006 & 2010 \\
United Kingdom & 39 & 33 & 27 & 6 & High & 0.255 & 0.028 & 1998 & 2025 \\
Qatar & 1 & 45 & 28 & 17 & High & 0.777 & 0.058 & 2008 & 2021 \\
Kuwait & 1 & 77 & 29 & 48 & Medium & 0.165 & 0.203 & 2009 & 2013 \\
Lithuania & 2 & 19 & 30 & -11 & Medium & 0.196 & 0.032 & 1998 & 2021 \\
Austria & 4 & 11 & 31 & -20 & Medium & 0.177 & 0.045 & 2000 & 2017 \\
Singapore & 1 & 34 & 32 & 2 & Medium & 0.198 & 0.050 & 2002 & 2017 \\
Bosnia and Herzegovina & 2 & 59 & 33 & 26 & Medium & 0.146 & 0.118 & 2006 & 2013 \\
Belgium & 5 & 15 & 34 & -19 & Medium & 0.178 & 0.048 & 2000 & 2016 \\
Ukraine & 5 & 60 & 35 & 25 & Medium & 0.239 & 0.058 & 2005 & 2018 \\
Thailand & 1 & 46 & 36 & 10 & Medium & 0.191 & 0.045 & 2002 & 2019 \\
Georgia & 1 & 50 & 37 & 13 & Medium & 0.209 & 0.050 & 2004 & 2019 \\
Czech Republic & 5 & 30 & 38 & -8 & Medium & 0.169 & 0.057 & 2003 & 2016 \\
New Zealand & 3 & 17 & 39 & -22 & Medium & 0.140 & 0.046 & 1999 & 2016 \\
Poland & 24 & 38 & 40 & -2 & Medium & 0.157 & 0.056 & 2002 & 2016 \\
Australia & 8 & 21 & 41 & -20 & Medium & 0.149 & 0.057 & 2002 & 2015 \\
Japan & 39 & 26 & 42 & -16 & Medium & 0.142 & 0.082 & 2003 & 2013 \\
Greece & 3 & 47 & 43 & 4 & Medium & 0.170 & 0.057 & 2004 & 2017 \\
Chile & 2 & 27 & 44 & -17 & Medium & 0.119 & 0.085 & 2003 & 2012 \\
Malaysia & 1 & 40 & 45 & -5 & Medium & 0.167 & 0.048 & 2003 & 2019 \\
New Caledonia & 1 & 32 & 46 & -14 & Medium & 0.563 & 0.025 & 2002 & 2032 \\
Morocco & 1 & 39 & 47 & -8 & Medium & 0.578 & 0.043 & 2007 & 2024 \\
Sri Lanka & 1 & 54 & 48 & 6 & Medium & 0.362 & 0.046 & 2006 & 2022 \\
Panama & 1 & 49 & 49 & 0 & Medium & 0.165 & 0.043 & 2002 & 2020 \\
Canada & 27 & 13 & 50 & -37 & Medium & 0.113 & 0.045 & 1998 & 2015 \\
Argentina & 3 & 64 & 51 & 13 & Medium & 0.183 & 0.075 & 2007 & 2017 \\
Hong Kong & 1 & 29 & 52 & -23 & Medium & 0.515 & 0.021 & 2000 & 2036 \\
Jordan & 1 & 62 & 53 & 9 & Medium & 0.131 & 0.075 & 2005 & 2016 \\
Costa Rica & 1 & 31 & 54 & -23 & Medium & 0.127 & 0.047 & 2000 & 2017 \\
El Salvador & 1 & 48 & 55 & -7 & Medium & 0.178 & 0.041 & 2002 & 2021 \\
Armenia & 1 & 83 & 56 & 27 & Medium & 0.151 & 0.104 & 2008 & 2016 \\
China & 2 & 70 & 57 & 13 & Medium & 0.170 & 0.085 & 2008 & 2017 \\
Mexico & 6 & 63 & 58 & 5 & Medium & 0.095 & 0.067 & 2003 & 2015 \\
Belarus & 1 & 71 & 59 & 12 & Medium & 0.189 & 0.059 & 2006 & 2019 \\
Iran & 1 & 85 & 60 & 25 & Medium & 0.462 & 0.098 & 2010 & 2018 \\
Tunisia & 1 & 72 & 61 & 11 & Medium & 0.105 & 0.084 & 2006 & 2016 \\
United Arab Emirates & 1 & 35 & 62 & -27 & Medium & 0.102 & 0.035 & 1998 & 2020 \\
Brazil & 28 & 52 & 63 & -11 & Medium & 0.128 & 0.044 & 2003 & 2020 \\
Russia & 42 & 58 & 64 & -6 & Medium & 0.117 & 0.054 & 2004 & 2018 \\
Namibia & 1 & 57 & 65 & -8 & Medium & 0.423 & 0.036 & 2005 & 2026 \\
Kazakhstan & 2 & 67 & 66 & 1 & Low & 0.118 & 0.060 & 2005 & 2018 \\
Venezuela & 2 & 44 & 67 & -23 & Low & 0.072 & 0.079 & 2003 & 2013 \\
Turkey & 10 & 55 & 68 & -13 & Low & 0.480 & 0.031 & 2005 & 2030 \\
Israel & 1 & 24 & 69 & -45 & Low & 0.083 & 0.120 & 2004 & 2010 \\
Ecuador & 2 & 61 & 70 & -9 & Low & 0.396 & 0.033 & 2005 & 2028 \\
South Africa & 3 & 56 & 71 & -15 & Low & 0.530 & 0.040 & 2007 & 2026 \\
Peru & 1 & 53 & 72 & -19 & Low & 0.088 & 0.044 & 2002 & 2019 \\
Jamaica & 1 & 65 & 73 & -8 & Low & 0.068 & 0.082 & 2004 & 2013 \\
Colombia & 4 & 69 & 74 & -5 & Low & 0.087 & 0.047 & 2003 & 2019 \\
Nepal & 1 & 84 & 75 & 9 & Low & 0.110 & 0.072 & 2007 & 2018 \\
Dominican Republic & 1 & 68 & 76 & -8 & Low & 0.052 & 0.136 & 2006 & 2011 \\
Paraguay & 1 & 87 & 77 & 10 & Low & 0.403 & 0.051 & 2008 & 2023 \\
Vietnam & 1 & 97 & 78 & 19 & Low & 0.365 & 0.050 & 2008 & 2023 \\
Indonesia & 4 & 81 & 79 & 2 & Low & 0.061 & 0.073 & 2006 & 2017 \\
Azerbaijan & 1 & 88 & 80 & 8 & Low & 0.342 & 0.055 & 2009 & 2022 \\
Kyrgyzstan & 1 & 74 & 81 & -7 & Low & 0.412 & 0.041 & 2008 & 2026 \\
Nicaragua & 1 & 66 & 82 & -16 & Low & 0.428 & 0.036 & 2007 & 2029 \\
Egypt & 1 & 82 & 83 & -1 & Low & 0.048 & 0.056 & 2004 & 2018 \\
India & 8 & 86 & 84 & 2 & Low & 0.394 & 0.048 & 2009 & 2025 \\
Algeria & 1 & 92 & 85 & 7 & Low & 0.404 & 0.043 & 2008 & 2026 \\
Albania & 1 & 80 & 86 & -6 & Low & 0.037 & 0.069 & 2005 & 2016 \\
Honduras & 1 & 79 & 87 & -8 & Low & 0.023 & 0.054 & 2004 & 2018 \\
Bolivia & 3 & 78 & 88 & -10 & Low & 0.016 & 0.059 & 2003 & 2016 \\
Lebanon & 1 & 75 & 89 & -14 & Low & 0.017 & 0.058 & 2003 & 2016 \\
Cambodia & 1 & 91 & 90 & 1 & Low & 0.015 & 0.065 & 2004 & 2016 \\
Uzbekistan & 1 & 89 & 91 & -2 & Low & 0.014 & 0.066 & 2004 & 2015 \\
Kenya & 1 & 73 & 92 & -19 & Low & 0.014 & 0.058 & 2003 & 2016 \\
Ghana & 1 & 94 & 93 & 1 & Low & 0.445 & 0.043 & 2010 & 2028 \\
Pakistan & 2 & 95 & 94 & 1 & Low & 0.010 & 0.066 & 2005 & 2017 \\
Nigeria & 1 & 99 & 95 & 4 & Low & 0.407 & 0.061 & 2012 & 2024 \\
Zimbabwe & 1 & 90 & 96 & -6 & Low & 0.006 & 0.066 & 2003 & 2015 \\
Uganda & 1 & 93 & 97 & -4 & Low & 0.415 & 0.061 & 2012 & 2024 \\
Bangladesh & 1 & 98 & 98 & 0 & Low & 0.403 & 0.069 & 2012 & 2023 \\
Cote d'Ivoire & 1 & 96 & 99 & -3 & Low & 0.003 & 0.072 & 2003 & 2014 \\
Angola & 1 & 100 & 100 & 1 & Low & 0.403 & 0.070 & 2012 & 2023 \\
\noalign{\hrule height 1.5pt}
\end{longtable}

%% file: feature_importance_tab.tex
\begin{tabular}{llr}
\noalign{\hrule height 1.5pt}
Rank & Feature & \multicolumn{1}{l}{Percent} \\ \hline
1 & Fraction online first difference Monday & 10.506 \\
2 & Fraction online Saturday & 8.656 \\
3 & Fraction online Monday & 6.785 \\
4 & Fraction online Tuesday & 6.361 \\
5 & Fraction online Friday & 6.147 \\
6 & Fraction online first difference Tuesday & 4.968 \\
7 & Fraction online first difference Saturday & 4.879 \\
8 & Fraction online first difference Thursday & 4.548 \\
9 & Fraction online first difference Friday & 4.468 \\
10 & Fraction online Wednesday & 4.346 \\
11 & Fraction online first difference Sunday & 4.130 \\
12 & Fraction online Thursday & 3.968 \\
13 & Fraction online first difference Wednesday & 3.638 \\
14 & Fraction online second difference Sunday & 3.483 \\
15 & Fraction online Sunday & 2.560 \\
16 & Fraction online second difference Saturday & 2.110 \\
17 & Fraction online second difference Friday & 2.021 \\
18 & Fraction online second difference Tuesday & 1.893 \\
19 & Fraction online second difference Thursday & 1.564 \\
20 & Wavelet coefficient 8 & 1.221 \\
21 & Fraction online second difference Wednesday & 1.180 \\
22 & Wavelet coefficient 10 & 1.084 \\
23 & Wavelet coefficient 5 & 1.027 \\
24 & Wavelet coefficient 6 & 0.962 \\
25 & Wavelet coefficient 7 & 0.961 \\
26 & Wavelet coefficient 9 & 0.959 \\
27 & Wavelet coefficient 2 & 0.898 \\
28 & Wavelet coefficient 3 & 0.843 \\
29 & Wavelet coefficient 1 & 0.833 \\
30 & Wavelet coefficient 4 & 0.815 \\
31 & Fraction online second difference Monday & 0.766 \\
32 & Latitude & 0.490 \\
33 & Dummy year 2007 & 0.343 \\
34 & Dummy year 2012 & 0.203 \\
35 & Dummy year 2011 & 0.113 \\
36 & Dummy year 2010 & 0.072 \\
37 & Dummy year 2009 & 0.055 \\
38 & Dummy year 2008 & 0.054 \\
39 & Trough Monday & 0.018 \\
40 & Trough Saturday & 0.014 \\
41 & Trough Sunday & 0.012 \\
42 & Trough Friday & 0.010 \\
43 & Trough Wednesday & 0.010 \\
44 & Trough Tuesday & 0.008 \\
45 & Trough Thursday & 0.008 \\
46 & Peak Thursday & 0.002 \\
47 & Peak Wednesday & 0.002 \\
48 & Peak Saturday & 0.001 \\
49 & Peak Monday & 0.001 \\
50 & Peak Sunday & 0.001 \\
51 & Peak Tuesday & 0.001 \\
52 & Peak Friday & 0.001 \\
\noalign{\hrule height 1.5pt}
\end{tabular}

%% file: sleep_tab.tex
\begin{table}[htpb]
\centering
\caption{Average predicted sleep start, sleep stop and sleep duration over the period 2006-2012 across cities and years by country}
\label{sleep-tab}
\begin{tabular}{lrrrr}
\noalign{\hrule height 1.5pt}
Country & \# City Years &  Sleep start & Sleep stop & Sleep duration \\ \hline
Japan & 106 & 23.06 & 6.33 & 7.27 \\
Slovenia & 4 & 22.58 & 6.03 & 7.46 \\
Suriname & 3 & 22.33 & 6.24 & 7.91 \\
Poland & 7 & 22.48 & 6.47 & 7.99 \\
New Zealand & 11 & 21.86 & 6.08 & 8.21 \\
Croatia & 4 & 22.45 & 6.70 & 8.25 \\
Indonesia & 5 & 22.18 & 6.48 & 8.30 \\
South Africa & 12 & 21.86 & 6.19 & 8.33 \\
Sweden & 27 & 22.48 & 6.91 & 8.43 \\
United States & 644 & 22.11 & 6.62 & 8.51 \\
Lebanon & 3 & 22.64 & 7.16 & 8.53 \\
Canada & 96 & 22.06 & 6.62 & 8.57 \\
United Kingdom & 86 & 21.98 & 6.64 & 8.66 \\
China & 12 & 22.38 & 7.04 & 8.66 \\
Denmark & 20 & 22.09 & 6.77 & 8.68 \\
Australia & 25 & 21.77 & 6.45 & 8.68 \\
Austria & 15 & 21.89 & 6.60 & 8.72 \\
Bulgaria & 10 & 22.51 & 7.25 & 8.74 \\
Romania & 6 & 22.31 & 7.11 & 8.80 \\
Israel & 8 & 22.12 & 6.96 & 8.85 \\
Brazil & 25 & 22.13 & 6.98 & 8.85 \\
Lithuania & 4 & 21.83 & 6.71 & 8.88 \\
Netherlands & 51 & 22.26 & 7.16 & 8.89 \\
Czech Republic & 15 & 21.79 & 6.69 & 8.91 \\
Venezuela & 4 & 21.52 & 6.43 & 8.91 \\
France & 130 & 22.13 & 7.07 & 8.93 \\
South Korea & 36 & 22.24 & 7.20 & 8.95 \\
Ireland & 5 & 22.59 & 7.56 & 8.97 \\
Puerto Rico & 5 & 22.50 & 7.46 & 8.97 \\
Norway & 9 & 22.53 & 7.55 & 9.02 \\
Germany & 77 & 21.86 & 6.94 & 9.07 \\
Chile & 3 & 21.74 & 6.86 & 9.13 \\
Finland & 10 & 21.78 & 6.94 & 9.16 \\
Greece & 6 & 22.56 & 7.74 & 9.19 \\
Switzerland & 10 & 21.80 & 7.03 & 9.23 \\
Portugal & 4 & 22.73 & 8.04 & 9.31 \\
Russia & 8 & 22.57 & 7.90 & 9.33 \\
Italy & 23 & 21.80 & 7.26 & 9.46 \\
Saudi Arabia & 3 & 20.58 & 6.07 & 9.48 \\
Belgium & 7 & 21.80 & 7.30 & 9.50 \\
Mexico & 42 & 21.85 & 7.61 & 9.75 \\
Turkey & 10 & 21.95 & 7.72 & 9.77 \\
Ukraine & 5 & 22.38 & 8.21 & 9.83 \\
Spain & 14 & 21.90 & 7.85 & 9.96 \\
Argentina & 3 & 22.48 & 8.74 & 10.26 \\ 
\noalign{\hrule height 1.5pt}
\end{tabular}
\end{table}